\documentclass[a4paper,12pt]{article}
\usepackage{epsfig}
\usepackage{amsmath}
\usepackage{amssymb}
\usepackage{amsfonts}
\usepackage{axodraw}
\usepackage{fleqn}
\usepackage{graphicx}
\usepackage{mathrsfs}
\usepackage[center,footnotesize,hang]{subfigure}

\def\be{\begin{equation}}
\def\ee{\end{equation}}
\def\beq{\begin{equation}}
\def\eeq{\end{equation}}
\def\bea{\begin{eqnarray}}
\def\eea{\end{eqnarray}}

\def\<{\left\langle}
\def\>{\right\rangle}

\def\SB{\tilde{S}/\tilde{B}'}
\def\tanb{\tan\beta}
\def\neut{\tilde{\chi}_1^0}

\def\s2w{\sin 2\theta_W}

\def\p1{p_1^\mu}
\def\p2{p_2^\mu}
\def\p3{p_3^\mu}

\DeclareMathOperator{\re}{Re}
\DeclareMathOperator{\im}{Im}

\def\nicefrac#1#2{\hbox{$\frac{#1}{#2}$}}

\allowdisplaybreaks[2]
\begin{document}
\bibliographystyle{OurBibTeX}
\begin{titlepage}
 \vspace*{-15mm}
\begin{flushright}
CERN-PH-TH/2008-219
\end{flushright}
\vspace*{5mm}
\begin{center}
{ \sffamily \LARGE Neutralino Dark Matter in the USSM}
\\[8mm]
J. Kalinowski$^{1,2}$, S.F. King$^3$ and J.P. Roberts$^{1,4}$
\\[3mm]
{\small\it
$^1$Physics Department, University of Warsaw, 00-681 Warsaw, Poland
\\[3mm]
$^2$Theory Division, CERN, CH-1211 Geneva 23, Switzerland
\\[3mm]
$^3$School of Physics and Astronomy,
University of Southampton,\\
Southampton, SO17 1BJ, U.K.
\\[3mm]
$^4$Center for Cosmology and Particle Physics,
New York University,\\
New York, NY 10003, USA
}\\[1mm]
\end{center}
\vspace*{0.75cm}
\begin{abstract}
\noindent This paper provides a comprehensive discussion of neutralino
dark matter within classes of extended supersymmetric models referred
to as the USSM containing one additional SM singlet Higgs
plus an extra $Z'$, together with their superpartners the singlino and
bino'. These extra states of the USSM can significantly modify the
nature and properties of neutralino dark matter relative to that of
the minimal (or even next-to-minimal) supersymmetric standard
models. We derive the Feynman rules for the USSM and calculate the
dark matter relic abundance and direct detection rates for elastic
scattering in the USSM for interesting regions of parameter space
where the largest differences are expected.

\end{abstract}
\end{titlepage}
\newpage
\setcounter{footnote}{0}

\section{Introduction}
One of the benefits of weak scale supersymmetry (SUSY) with
conserved R parity is that the lightest supersymmetric particle
(LSP) is absolutely stable, and provides a weakly interacting
massive particle (WIMP) candidate
\cite{EHNOS,hep-ph/9506380} capable of accounting for the
observed cold dark matter (CDM) relic density $\Omega_{CDM} h^2=
0.1099 \pm 0.0062$ \cite{wmap5}.  In particular, the lightest
neutralino in SUSY models is an excellent candidate, providing its
mass, composition and interactions are suitably tuned to result in
the correct value of $\Omega_{CDM} h^2$. The minimal
supersymmetric standard model (MSSM) has become a widely studied
paradigm \cite{Chung:2003fi}. However the stringent upper bound on
the Higgs boson mass in the MSSM combined with its experimental
lower bound from LEP has led to some tension in the electroweak
symmetry breaking sector, roughly characterized by a fine-tuning
of parameters at the percent level \cite{Kane:1998im}.  While the
experimental elusiveness of the Higgs boson may cast some doubt on
the MSSM, there are a host of non-minimal SUSY models which
predict a heavier and/or more weakly coupled Higgs boson
\cite{Chung:2003fi}.

A further reason to move beyond the minimal case is the so-called mu
problem of the MSSM \cite{Cohen:2008ni}. The MSSM
contains a bilinear mass term that couples the two Higgs doublets with
a dimensionful coupling $\mu$. This term is SUSY preserving, and as
such only has two natural values, $\mu=0$ and $\mu=M_{Pl}$ (unless
special forms of the Kahler metric are assumed).  However experimental
data and the stability of the Higgs mass requires that $\mu$ be of the
order of the SUSY breaking scale. In non-minimal SUSY models the mu
problem is solved by setting $\mu=0$ and including an additional
superfield $\hat S$, a singlet under the Standard Model (SM) gauge
group, which couples to the Higgs doublet superfields $\hat H_1, \hat
H_2$ according to $ \lambda \hat S\hat H_1\hat H_2 $, where $\lambda$
is a dimensionless coupling constant.  We shall refer to such models
generically as singlet SUSY models.  Such a coupling replaces the SUSY
Higgs/Higgsino mass term $\mu \hat H_1\hat H_2$ of the MSSM.  The
singlet vacuum expectation value (VeV) $\langle S\rangle$ then
dynamically generates a SUSY Higgs/Higgsino mass near the weak scale
as required. This results in an increased Higgs boson mass upper bound
depending on the value of $\lambda$, and hence a welcome reduction in
electroweak fine tuning in addition to solving the $\mu$ problem of
the MSSM \cite{BasteroGil:2000bw}.

However, although an extra singlet superfield $\hat S$ seems like a
minor modification to the MSSM, which does no harm to either gauge
coupling unification or neutralino dark matter, there are further
costs involved in this scenario since the introduction of the singlet
superfield $\hat S$ leads to an additional accidental global $U(1)_X$
(Peccei-Quinn (PQ) \cite{Peccei:1977hh}) symmetry which will result in a weak scale massless
axion when it is spontaneously broken by $\langle S\rangle$ \cite{Fayet:1974fj}. Since
such an axion has not been observed experimentally, it must be removed
somehow. This can be done in several ways resulting in different
non-minimal SUSY models, each involving additional fields and/or
parameters. For example, the classic solution to this problem is to
introduce a singlet term $\hat S^3$, as in the next-to-minimal
supersymmetric standard model (NMSSM)~\cite{nmssm}, which reduces the
PQ symmetry to the discrete symmetry $Z_3$. The subsequent breaking of
a discrete symmetry at the weak scale can lead to cosmological domain
walls which would overclose the Universe. This can be avoided by
breaking the $Z_3$ symmetry explicitly without upsetting the hierarchy
problem by non-renormalizable operators that obey a $Z_2$
$R$-symmetry~\cite{Panagiotakopoulos:1998yw}, or by removing the $\hat
S^3$ term altogether~\cite{Panagiotakopoulos:1999ah}.

Another solution to the axion problem of singlet models, which we
follow, is to promote the PQ symmetry to an Abelian $U(1)_X$ gauge
symmetry \cite{Fayet:1977yc}. The idea is that the extra gauge boson will eat the
troublesome axion via the Higgs mechanism resulting in a massive $Z'$
at the TeV scale. The essential additional elements of such a scenario
then consist of two extra superfields relative to those of the MSSM,
namely the singlet superfield $\hat S$ and the $U(1)_X$ gauge
superfield $B'$.  The scenario involving only the MSSM superfields
plus these two additional superfields, may be considered as a
phenomenological model in its own right which has been referred to as
the USSM. In the USSM, then, the MSSM particle spectrum is extended by
a new CP-even Higgs boson $S$, a gauge boson $Z'$ and two neutral
--inos: a singlino $\tilde{S}$ and a bino' $\tilde{B}'$ while other
sectors are not enlarged. The presence of new singlino and bino'
states greatly modifies the phenomenology of the neutralino sector
both at colliders and in cosmology-related processes. The collider
phenomenology and cosmology of the USSM has been studied in
\cite{ussm,deCarlos:1997yv,Cvetic:1997ky,u1extensions,Choi:2006fz,Barger:2007nv,our},
which we briefly review as follows.

The collider phenomenology of the USSM has recently been considered in
\cite{Choi:2006fz}.  The neutralino production cross sections and
their decay branching fractions depend crucially on their masses and
composition with respect to the MSSM case. If the new -ino states are
heavy, their influence on the MSSM neutralinos is small. In contrast,
if the singlino mass scale is low, the production rates can be quite
different and, since there are more neutralinos, the decay chains of
sparticles can be longer.  Moreover, if the mass gaps between the MSSM
and new -inos are very small, the standard decay modes are almost shut
and radiative transitions between neutralino states with a soft photon
may be dominant. In such a case the decay chains can be apparently
shorter, a feature which is of relevance for the LHC experiments.

The dark matter phenomenology of the USSM was first studied in
\cite{deCarlos:1997yv,Cvetic:1997ky}, and more recently in
\cite{Barger:2007nv}. In \cite{deCarlos:1997yv}, the analysis was
performed in a scenario with a very light $Z'$ and considered the case
in which the LSP was a very light singlino. This allowed the authors
to consider the annihilation of dark matter in the early universe to
be dominated by s-channel $Z'$ processes, allowing an analytic
solution to the dark matter relic density to be obtained.  In the full
parameter space of the USSM, this is just one possibility, and indeed
such a light $Z'$ is heavily disfavored by current data. In
\cite{Barger:2007nv} the recoil detection of the dark matter candidate
in the USSM (and other non-minimal SUSY models) was considered.

In this paper, we provide an up to date and comprehensive analysis of
neutralino dark matter in the USSM\footnote{The recent observation of
a positron excess by the PAMELA collaboration \cite{pamela} have
caused a flurry of speculation that the high energy positrons are
produced by annihilating dark matter in the galactic
halo\cite{positronDM}. An alternative explanation is that
astrophysical sources could account for the positron excess - in
particular nearby pulsars \cite{pulsars}. It is unclear as yet which
of these explanations is correct and as a result we do not address the
PAMELA results further in this work.}. We provide a complete
discussion of the extended gauge, neutralino, Higgs squark and slepton
sectors in the USSM, and using the LanHEP \cite{lanhep} package,
derive all the new Feynman rules involving these extended sectors. We
first provide a complete qualitative discussion of the new
annihilation channels relevant for the calculation of the cold dark
matter relic density for the neutralino LSP in the USSM. We also
discuss the elastic scattering cross section for the neutralino LSP in
the USSM, including both spin-independent and spin-dependent parts of
the cross sections, relevant for the direct dark matter search
experiments. We then survey the parameter space of the USSM, and
discuss quantitatively how the nature and composition of the
neutralino LSP can be significantly altered compared to that in the
MSSM due to the extra singlino and bino' states, for different ranges
of parameters. The Feynman rules are then implemented into the
micrOMEGAs \cite{micromegas} package in order to calculate the relic
density for the corresponding regions of parameter space. This
provides a full calculation of the annihilation channels including
co-annihilation and careful treatment of resonances as well as
accurately calculating the relic density for an arbitrary admixture of
states. In this way we extend the analysis of USSM neutralino dark
matter annihilation beyond the specific cases previously studied in
the literature. We also perform an equally general calculation of the
direct detection cross-sections for USSM dark matter for elastic
neutralino--nuclei scattering.

It is worth emphasizing that the USSM is not a complete model, since
from its definition it does not include the additional superfields at
the TeV scale, charged under the gauged Abelian symmetry, which are
necessarily present in order to cancel the fermionic gauge anomalies
involving the $U(1)_X$ gauge symmetry.  For example, a well motivated
and elegant solution to the problem of anomaly cancelation is to
identify the Abelian gauge group as a subgroup of $E_6$ and then
cancel the anomalies by assuming complete 27 dimensional
representations of matter down to the TeV scale.  With the further
requirement that the right-handed neutrino carries zero charge under
the Abelian gauge group (in order to have a high scale see-saw
mechanism) this then specifies the theory uniquely as the
E$_6$SSM~\cite{Keith:1997zb,King:2005jy}.
However our working assumption
is that the additional matter superfields required to cancel anomalies
are heavy compared to the $Z'$ mass. The USSM considered in this paper
may thus be regarded as a low-energy truncation of the E$_6$SSM
model, with other E$_6$SSM fields assumed heavy,
and the charge assignments under the extra $U(1)_X$ as given in
\cite{King:2005jy} and summarized in Section 2.

Despite that fact that the USSM must be regarded as a truncation of a
complete model, it makes sense to study the physics and cosmology in
the USSM since it provides a simplified setting to learn about crucial
features which will be relevant to any complete model involving an
additional $U(1)_X$ gauge group and a singlet. For example, as already
mentioned the neutralino LSP in the USSM may have components of the
extra gaugino $\tilde{B}'$ and singlino $\tilde{S}$, in addition to
the usual MSSM neutralino states. Naively we might expect that the
dark matter phenomenology of such regions would be similar to that of
singlino dark matter in the NMSSM. However this is not the case. The
inclusion of the bino' state, as well as the lack of a cubic
interaction term $\hat S^3$, results in a significant change in the
phenomenology. Also the neutralino mass spectrum in the USSM is very
different from that of the NMSSM as the singlino mass is determined
indirectly by a mini-see-saw mechanism involving the bino' soft mass
parameter $M_1'$ rather than through a diagonal mass term arising from
the cubic $\hat S^3$. The lack of a cubic interaction term also
restricts the annihilation modes of the singlino, making it dominantly
reliant on annihilations involving non-singlet Higgs bosons and
higgsinos. As the USSM has a different Higgs spectrum to the NMSSM,
notably in the pseudoscalar Higgs sector, the Higgs dominated
annihilation channels of the USSM singlino are significantly modified
with respect to the NMSSM singlino. As Higgs exchange diagrams
dominate the direct detection phenomenology, the difference in the
Higgs spectrum and the singlino interactions results in significant
differences in the direct detection predictions as well.

The remainder of the paper is organized as follows.  In section 2 we
shall define the Lagrangian of the USSM and discuss the Higgs, $Z'$,
neutralino and sfermion sectors. In sections 3 and 4 we shall give an
overview of the important features of the relic density calculation
and the direct detection calculation, respectively, highlighting the
main differences to the MSSM. In section 5 we present the results of
the full numerical calculations for both the relic density and the
direct detection cross-section. It will be performed in two physically
interesting scenarios: (A) with the MSSM higgsino and gaugino mass
parameters fixed, while the mass of the extra U(1) gaugino taken free
(to complement the collider phenomenology discussed in
Ref.~\cite{Choi:2006fz}); (B) with GUT-unified gaugino masses. Section
6 summarizes and concludes the paper. The mass matrix structure of the
extended Higgs scalar sector, and a discussion of the Feynman rules
involving the extended neutralino sector in the USSM are given in a
pair of appendices.

\section{The USSM model}

Including the extra U(1)$_X$ symmetry, the gauge group of the model is
$G={\rm SU}(3)_C\times{\rm SU}(2)_L\times{\rm U}(1)_Y\times{\rm
U}(1)_X$ with the couplings $g_3, g_2, g_Y, g_X$, respectively. In
addition to the MSSM superfields, the model includes a new vector
superfield $\hat{B}_X$ and a new iso--singlet Higgs superfield
$\hat{S}$. The usual MSSM Yukawa terms $\hat{W}_Y$ of the MSSM
superpotential ({\it i.e.} without the $\mu$ term) are augmented by an
additional term that couples the iso--singlet to the two iso--doublet
Higgs fields:
\begin{eqnarray}\label{eqi}
\hat{W}=\hat{W}_Y +\lambda \hat{S}\, (\hat{H}_u\hat{H}_d)\,.
\label{eq:superpotential}
\end{eqnarray}
The coupling $\lambda$ is dimensionless. Gauge invariance of the
superpotential $\hat{W}$ under U(1)$_X$ requires the U(1)$_X$ charges
to satisfy $Q^X_{H_d}+Q^X_{H_u}+Q^X_S=0$ and corresponding relations
between the U(1)$_X$ charges of Higgs and matter fields. In the
following we adopt the U(1)$_X$ charges as in the E$_6$SSM
model~\cite{King:2005jy}, see Table \ref{tab:charges}. (For notational
convenience we will also use $Q_1=Q^X_{H_d}$, $Q_2=Q^X_{H_u}$ and
$Q_S=Q^X_S$.) The effective $\mu$ parameter is generated by the vacuum
expectation value $\langle S\rangle$ of the scalar $S$--field.

\begin{table}[ht]
  \centering
  \begin{tabular}{|c|c|c|c|c|c|c|c|c|c|}
    \hline
 $i$ & $Q$ & $u^c$ & $d^c$ & $L$ & $e^c$ & $N^c$ & $S$ & $H_2$ & $H_1$  \\
 \hline
$\sqrt{\frac{5}{3}}Q^{Y}_i$
 & $\frac{1}{6}$ & $-\frac{2}{3}$ & $\frac{1}{3}$ & $-\frac{1}{2}$
& $1$ & $0$ & $0$ & $\frac{1}{2}$ & $-\frac{1}{2}$  \\
 \hline
$\sqrt{{40}}Q^{X}_i$
 & $1$ & $1$ & $2$ & $2$ & $1$ & $0$ & $5$ & $-2$ & $-3$  \\
 \hline
  \end{tabular}
  \caption{\it\small The $U(1)_Y$ and $U(1)_{X}$ charges of matter
    fields in the USSM, where $Q^{X}_i$ and $Q^{Y}_i$ are defined with
    the correct $E_6$ normalization factor required for the RG
    analysis \cite{King:2005jy}.}
  \label{tab:charges}
\end{table}

The USSM particle content, in addition to the MSSM particles, includes
a single extra scalar state, a new Abelian gauge boson and an
additional neutral higgsino and gaugino state. The chargino sector
remains unaltered, while the sfermion scalar potential receives
additional D-terms.

\subsection{The Abelian gauge sector}\label{sec:kin-mix}

With two Abelian gauge factors, U(1)$_Y$ and U(1)$_X$, the two sectors
can mix through the coupling of the kinetic parts \cite{holdom},
\begin{eqnarray} \label{lgauge}
{\cal L}_{\rm gauge}
 = \frac{1}{32}\int d^2\theta
   \left\{\hat{W}_Y \hat{W}_Y + \hat{W}_X \hat{W}_X
        +2\sin\chi\, \hat{W}_Y \hat{W}_X\right\}\,,
\end{eqnarray}
where $\hat{W}_Y$ and $\hat{W}_X$ are the corresponding chiral
superfields.  The gauge/gaugino part of the Lagrangian can be
converted back to the canonical form by the GL(2,$\mathbb{R}$)
transformation from the original superfield basis $\hat{W}_Y,\,
\hat{W}_X$ to the new one $\hat{W}_B,\,\hat{W}_{B'}$
~\cite{holdom,kin_mix}:
\begin{eqnarray} \label{kinmixmatrix}
\left(\begin{array}{c} \hat{W}_Y \\ \hat{W}_X
      \end{array}\right)
 =
\left(\begin{array}{cc}
         1   & -\tan\chi \\
         0   & 1/\cos\chi
      \end{array}\right)
\left(\begin{array}{l}
        \hat{W}_B \\
        \hat{W}_{B'}
       \end{array}\right)\,.
\label{eq:trans}
\end{eqnarray}
This transformation alters the U(1)$_Y\times$U(1)$_X$ part of the
covariant derivative to
\begin{eqnarray}
D_\mu &=&\;\partial_\mu + i g_Y Y_i B_\mu
   +i (- g_Y Y_i \tan\chi +\frac{g_X}{\cos\chi} Q^X_i ) B'_\mu  \\
    &=&\;\partial_\mu + i g_1 Y_i B_\mu +i g'_1 Q'_i B'_\mu\,,
\end{eqnarray}
where we introduced the notation $g_1=g_Y$, $g'_1=g_X/\cos\chi$.  We
will also use $g'=g_1\sqrt{3/5}$ for the low-energy (non-GUT
normalized) hypercharge gauge coupling.

With the above mixing matrix the hypercharge sector of the Standard
Model is left unaltered, while the effective U(1)$_X$ charge is
shifted from its original value $Q_i^X$ to
\begin{eqnarray}
Q'_i=Q^X_i - \frac{g_1}{g'_1} Y_i \tan\chi\,.\label{eq:effQ}
\end{eqnarray}
As a result of the kinetic mixing, new interactions among the gauge
bosons and matter fields are generated even for matter fields with
zero U(1)$_X$ charge.

In the E$_6$SSM the two U(1) gauge groups are automatically orthogonal
at the GUT scale and the RG running effects give acceptable small
mixing at the low scale~\cite{King:2005jy,Langacker:2008yv} providing in a natural way
the general agreement between SM analyses and precision
data~\cite{Zprime_bound}. Therefore in the reminder of the paper we
will simply write $Q_i$ instead of $Q'_i$.

After breaking the electroweak and U(1)$_X$ symmetries spontaneously
due to non--zero vacuum expectation values of the iso--doublet and the
iso--singlet Higgs fields,
\begin{eqnarray}
\langle H_u \rangle
  = \frac{\sin\beta}{\sqrt{2}}
    \left(\begin{array}{c}
              0 \\
              v
          \end{array}\right),\qquad
\langle H_d \rangle
  = \frac{\cos\beta}{\sqrt{2}}
    \left(\begin{array}{c}
              v \\
              0
          \end{array}\right),\qquad
\langle S \rangle = \frac{1}{\sqrt{2}} v_S\,,
\end{eqnarray}
the $Z,Z'$  mass matrix takes the form
\begin{equation}
M_{ZZ'}^{2}=\left(\begin{array}{cc}
M_{Z}^{2} & \Delta^{2}\\
\Delta^{2} & M_{Z'}^{2}\end{array}\right),\label{eq:Mzz'}\end{equation}
where
\begin{eqnarray}
M_{Z}^{2} & = & \frac{g'^{2}+g_{2}^{2}}{4}v^{2}\nonumber \\
\Delta^{2} & = & \frac{g'_{1}\sqrt{g'^{2}+g_{2}^{2}}}{2}v^{2}\left(Q_{1}\cos^{2}\beta-
Q_{2}\sin^{2}\beta\right)\nonumber \\
M_{Z'}^{2} & = & g_{1}'^{2}v^{2}\left(Q^2_{1}\cos^{2}\beta+Q^2_{2}\sin^{2}\beta\right)+
g_{1}'^{2}Q^2_{S}v_{S}^{2}\label{eq:MzMz'Delta}
\end{eqnarray}
We then diagonalise the mass matrix to give the mass eigenstates:
\begin{equation}
\left(\begin{array}{c}
Z_1\\Z_2
\end{array}\right)
=D_{ij}
\left(\begin{array}{c}
Z\\Z'
\end{array}\right)
=
\left(\begin{array}{cc}
\cos\alpha_{ZZ'} & \sin\alpha_{ZZ'}\\
-\sin\alpha_{ZZ'} & \cos\alpha_{ZZ'}
\end{array}\right)
\left(\begin{array}{c}
Z\\Z'
\end{array}\right)\label{eq:Z1Z2}
\end{equation}
where the resultant masses and $ZZ'$ mixing angle are given by
\begin{eqnarray}
&&M_{Z_{1},Z_{2}}^{2}=\frac{1}{2}\left(M_{Z}^{2}+M_{Z'}^{2}\mp\sqrt{\left(M_{Z}^{2}-
M_{Z'}^{2}\right)^{2}+4\Delta^{4}}\right)\label{eq:Mz1Mz2}\\
&&\alpha_{ZZ'}=\frac{1}{2}\arctan\left(\frac{2\Delta^{2}}{M_{Z'}^{2}-
M_{Z}^{2}}\right)\label{eq:alphaZZ'}
\end{eqnarray}

\subsection{The Higgs sector}

In the charged sector it is convenient to introduce the $G^\pm,H^\pm$
basis as:
\begin{eqnarray}
&&G^{-} = H_{d}^{-}\cos\beta-H_{u}^{+*}\sin\beta\label{chargedHiggs}\\
&&H^{+} = H_{d}^{-*}\sin\beta+H_{u}^{+}\cos\beta\nonumber
\end{eqnarray}
After the gauge symmetry breaking, two Goldstone modes $G^\pm$ from
the original $H_u$ and $H_d$ doublets are eaten by $W^\pm$ fields
leaving two physical charged Higgs bosons $H^\pm$, with the mass
\begin{equation}
m_{H^{\pm}}^{2}=\frac{\sqrt{2}\lambda
A_{\lambda}}{\sin2\beta}v_{S}-\frac{\lambda^{2}}{2}v^{2}
+\frac{g_{2}^{2}}{2}v^{2}+\Delta_{\pm},\label{mH+-}
\end{equation}
where the trilinear coupling $A_\lambda$ is the soft-SUSY breaking
counterpart of $\lambda$, and the one-loop corrections $\Delta_{\pm}$
are the same as in the MSSM~\cite{delta-pm} with the effective $\mu$
parameter given by
\begin{equation*}
\mu\equiv\lambda\frac{v_S}{\sqrt{2}}.
\end{equation*}

In the CP-conserving model the CP-even and CP-odd scalar Higgs
component fields do not mix. The CP-even sector involves $\re H^0_d$,
$\re H^0_u$ and $\re S$ fields. The $3\times 3$ mass matrix of the
CP-even Higgs scalars $M^2_{even}$ has been calculated to one-loop in
Refs.~\cite{King:2005jy,kovalenko} in the field space basis
$h,H,S$. This basis\footnote{Note that $h,H$ are not the MSSM-like
eigenstates.} is rotated by an angle $\beta$ with respect to the
interaction basis,
\begin{eqnarray}
\sqrt{2}\re \left(\begin{array}{c} H_{d}^{0}\\ H_{u}^{0}\\ S \end{array}\right)
= \left( \begin{array}{ccc} \cos\beta& -\sin\beta & 0\\
\sin\beta & \cos\beta & 0\\
0 & 0 & 1\end{array} \right) \left(\begin{array}{c}h\\ H\\ N\end{array}\right)
+\left(\begin{array}{c} v\cos\beta \\
v\sin\beta \\  v_{S}\end{array}\right)
\label{evenH}
\end{eqnarray}
The explicit form of $M^2_{even}$ is given in Appendix A.  It can be
diagonalized by a $3\times 3$ orthogonal mixing matrix $\cal(O)$, i.e.
\begin{eqnarray}
M^{2\, diag}_H={\cal{O}}^T M^2_{even}{\cal{O}}
\end{eqnarray}
by going to the mass eigenstates basis
\begin{eqnarray}
\label{hdef}(H_1,H_2,H_3)=(h,H,N)\cal{O}
\end{eqnarray}
in which, by convention, mass eigenstates are ordered by mass,
$m_{H_i}\leq m_{H_{i+1}}$.

It will be convenient to introduce a mixing matrix ${\cal O}'$,
\begin{eqnarray}\label{eq:defOp}
{\cal O}'=\left( \begin{array}{ccc} \cos\beta& -\sin\beta & 0\\
\sin\beta & \cos\beta & 0\\
0 & 0 & 1\end{array} \right){\cal O},
\end{eqnarray}
that enters the Feynman rules. It is a superposition of two rotations
in eqs.~(\ref{evenH}) and (\ref{hdef}) and links the interaction
eigenstates $H^0_d,H^0_u,S$ directly to the CP-even mass eigenstates
$H_1,H_2,H_3$.

The imaginary parts of the neutral components of the Higgs doublets
and Higgs singlet compose the CP-odd sector of the model. In the field
basis $A,G,G'$ defined by
\begin{eqnarray}
\sqrt{2}\im H_{d}^{0} & = & G\cos\beta+(A\cos\phi-G'\sin\phi)\sin\beta\nonumber \\
\sqrt{2}\im H_{u}^{0} & = & -G\sin\beta+(A\cos\phi-G'\sin\phi)\cos\beta\nonumber \\
\sqrt{2}\im S & = & A\sin\phi+G'\cos\phi\label{oddH}
\end{eqnarray}
the massless pseudoscalar $G,G'$ fields are absorbed to $Z,Z'$ after
the electroweak gauge symmetry breaking. The physical CP-odd Higgs
boson $A$ acquires mass
\begin{equation}
  m_{A}^{2}=\frac{\sqrt{2}\lambda A_{\lambda}}{\sin2\phi}v+\Delta_{EA}
  \label{m_A}
\end{equation}
where $\tan\phi=v\sin2\beta/2v_S$ and the one-loop correction
$\Delta_{EA}$ is given in Appendix A.

Note that the Higgs sector of this model involves only one physical
CP-odd pseudoscalar as in the MSSM, since, unlike the NMSSM, the extra
CP-odd state arising from the singlet is eaten by the $Z'$. However,
there are three CP-even scalars, one more than in the MSSM, where the
extra singlet state arises from the extra singlet as in the NMSSM.
The characteristic Higgs mass spectrum in this model is governed by
the value of $\lambda$.  For small values of $\lambda$, say $\lambda <
g_1$, the Higgs spectrum resembles that of the MSSM, with the heaviest
CP-even Higgs scalar being predominantly composed of the singlet
scalar state, and being approximately degenerate with the CP-odd
pseudoscalar and the charged Higgs states when their masses exceed
about 500 GeV. In this regime the lightest CP-even Higgs scalar is
Standard Model like, and respects the MSSM mass bound.  On the other
hand, for large values of $\lambda$, say $\lambda > g_1$, a viable
Higgs mass spectrum only occurs for a very large CP-odd Higgs mass,
say $m_A\approx 2-3$ TeV, with the heaviest CP-even Higgs scalar being
non-singlet and degenerate with the the CP-odd and charged Higgs
states. The second heaviest Higgs scalar is comprised mainly of the
singlet state and is thus unobservable, while the lightest CP-even
Higgs scalar is Standard Model like but may significantly exceed the
MSSM bound.  For more details concerning the Higgs sector see
\cite{King:2005jy}.

\subsection{The neutralino sector}

The Lagrangian of the neutralino system follows from the
superpotential in Eq.$\,$(\ref{eq:superpotential}), complemented by
the gaugino SU(2)$_L$, U(1)$_Y$ and U(1)$_X$ mass terms of the
soft--supersymmetry breaking electroweak Lagrangian:
\begin{eqnarray}
{\cal L}^{\rm gaugino}_{\rm mass}
 &=& -\frac{1}{2} M_2 \tilde{W}^a\tilde{W}^a
-\frac{1}{2} M_Y \tilde{Y} \tilde{Y}
   -\frac{1}{2} M_X \tilde{X} \tilde{X}
   - M_{YX} \tilde{Y}\tilde{X} + {\rm h.c.}
\end{eqnarray}
where the $\tilde{W}^a$ ($a=1,2,3$), $\tilde{Y}$ and $\tilde{X}$ are
the (two--component) SU(2)$_L$, U(1)$_Y$ and U(1)$_X$ gaugino fields,
and $M_i$ ($i=2,X,Y,YX$) are the corresponding soft-SUSY breaking mass
parameters.  After performing the transformation of gauge superfields
to the gauge boson eigenstate basis, Eq.(\ref{eq:trans}), the
Lagrangian takes the form
\begin{eqnarray} \label{gauginomass}
{\cal L}^{\rm gaugino}_{\rm mass}
 &=&  -\frac{1}{2} M_2 \tilde{W}^a\tilde{W}^a
-\frac{1}{2} M_1  \tilde{B}  \tilde{B}
     -\frac{1}{2} M'_1 \tilde{B}' \tilde{B}'
     -M_K \tilde{B}\tilde{B}' + {\rm h.c.}\,,
\end{eqnarray}
where
\begin{eqnarray} \label{Mfactors}
M'_1 \equiv \frac{M_X}{\cos^2\chi}-\frac{2\sin\chi}{\cos^2\chi} M_{YX}
+ M_Y \tan^2\chi\,,\qquad M_K\equiv \frac{M_{YX}}{\cos\chi}-M_Y
\tan\chi\,,
\end{eqnarray}
and we introduce the conventional notation for the U(1) bino mass
$M_1\equiv M_Y$.  In parallel to the gauge kinetic mixing discussed in
Sect.\ref{sec:kin-mix}, the Abelian gaugino mixing mass parameter
$M_{YX}$ is assumed small compared with the mass scales of the gaugino
and higgsino fields.

Notice that the gauge kinetic term mixing (and the corresponding
soft-SUSY breaking mass) can be a source of mass splitting between the
$\tilde{B}$ and $\tilde{B}'$ gauginos in models with universal gaugino
masses $M_X=M_Y$. Since the mixing angle $\chi$ must be small, as
required by data \cite{Zprime_bound}, the splitting is very small. The
splitting could be enhanced if additional U(1) gauge factors in the
hidden sector were present that mix via the kinetic term with the
visible sector.\footnote{ Since the fields in the hidden sector are
generally considered to be heavy enough and the hidden-visible mixing
is expected to be small, their effect on the visible gauge sector can
be negligible. Nevertheless, the mass of the Abelian gaugino in the
visible sector can obtain a substantial contribution, as advocated in
\cite{Suematsu:2006wh}.} In our phenomenological analyses, therefore,
we will consider two scenarios: (A) with $M'_1$ taken as a free
parameter, independent from $M_1$; and (B) with $M'_1$ tight to $M_1$
and $M_2$ by a unification of gaugino masses at the GUT scale.

After breaking the electroweak and U(1)$_X$ symmetries spontaneously
the doublet higgsino mass $\mu$ and the doublet higgsino--singlet
higgsino mixing $\mu_\lambda$ parameters are generated
\begin{equation}
\mu\equiv \lambda \frac{v_S}{\sqrt{2}} \;\;\;\;\; {\rm{and}}
\;\;\; \;\; \mu_\lambda\equiv\lambda \frac{v}{\sqrt{2}}\,.
\end{equation}
The USSM neutral gaugino--higgsino mass matrix in a basis of
two--component spinor fields $\xi\equiv(\tilde{B}, \tilde{W}^3,
\tilde{H}^0_d, \tilde{H}^0_u, \tilde{S}, \tilde{B}')^T$ can be written
in the following block matrix form
\begin{equation}
\!\!\!\!\!\!\!\!\!\!\!\!\!\!M_{\tilde{\chi}^0}=\left(\begin{array}{cccc|cc}
M_{1} & 0 & -M_Z\, c_\beta s_W & M_Z\, s_\beta s_W & 0 & M_K\\ 0 &
M_{2} & M_Z\, c_\beta c_W & -M_Z\, s\beta c_W & 0 & 0\\ -M_Z\, c_\beta
s_W & M_Z\, c_\beta c_W & 0 & -\mu & -\mu_\lambda\,s_\beta &
Q_{1}g'_{1}v c_\beta\\ M_Z\, s_\beta s_W & -M_Z\, s\beta c_W & -\mu &
0 & -\mu_\lambda\,c_\beta & Q_{2}g'_{1}v s_\beta\\[2mm] \cline{1-6}
\rule{0cm}{5mm} 0 & 0 & -\mu_\lambda\,s_\beta & -\mu_\lambda\,c_\beta
& 0 & Q_{S}g'_{1}v_{S}\\ M_K& 0 & Q_{1}g'_{1}v c_\beta &
Q_{2}g'_{1}v s_\beta & Q_{S}g'_{1}v_{S} &
M'_{1}\end{array}\right)\label{eq:general_mass_matrix}
\end{equation}
where the upper-left $4\times 4$ is the neutral gaugino--higgsino mass
matrix of the MSSM, the lower-right $2\times 2$ corresponds to the new
sector containing the singlet higgsino (singlino) and the new
U(1)--gaugino $\tilde B'$ that is orthogonal to the bino~$\tilde B$,
and off-diagonal $4\times 2$ describes the coupling of the two sectors
via the neutralino mass matrix ($s_\beta\equiv\sin\beta$,
$c_\beta\equiv\cos\beta$, and $s_W, c_W$ are the sine and cosine of
the electroweak mixing angle $\theta_W$). Notice the see-saw type
structure of the new sector due to the absence of a diagonal mass
parameter for the singlino $\tilde S$, which is in direct contrast to
the NMSSM in which the cubic self-interaction generates a singlet mass
term\cite{nmssm}. For the same reason, in the USSM the lightest
neutralino can never be bino'-dominated.

In general, the neutralino mass matrix $M_{\tilde{\chi}^0}$ is a
complex symmetric matrix.  To transform this matrix to the diagonal
form, we introduce a unitary $6\times 6$ matrix $N$ such that
\begin{eqnarray} \label{ndef}
\tilde{\chi}^0_k = N_{k\ell}\, ( \tilde{B}, \tilde{W}^3, \tilde{H}_d,
    \tilde{H}_u, \tilde{S}, \tilde{B}')_\ell\,,
\end{eqnarray}
where the physical neutralino states $\tilde{\chi}^0_k$ $[k =1,...,6]$
are ordered according to ascending absolute mass values. The
eigenvalues of the above matrix can be of both signs; the negative
signs are incorporated to the mixing matrix $N$.  Mathematically, this
procedure of transforming a general complex symmetric matrix to the
diagonal form with non-negative diagonal elements is called the Takagi
diagonalization, or the singular value
decomposition~\cite{Choi:2006fz, takagi}.  Physically, the unitary
matrix $N$ determines the couplings of the mass--eigenstates
$\tilde{\chi}^0_k$ to other particles.

Although the complexity of neutralino sector increases dramatically by
this extension as compared to the MSSM (which can be solved
analytically), the structure remains transparent since, in fact, the
original MSSM and the new degrees of freedom are coupled weakly.
$M_K$ must be small by the requirement that the mixing of the $U(1)_X$
and $U(1)_Y$ sectors satisfy experimental limits. The remaining
off-diagonal terms are suppressed with respect to the corresponding
block diagonal terms by a factor of $v/v_S$. Since $v_S$ sets the mass
of the $Z'$, this results in $v_S$ being roughly an order of magnitude
greater than $v$. Therefore in physically interesting case of weak
couplings of both the MSSM higgsino doublets to the singlet higgsino
and to the U(1)$_X$ gaugino, and the coupling of the U(1)$_Y$ and
U(1)$_X$ gaugino singlets, the remaining terms in the off-diagonal
$4\times 2$ submatrix in Eq.$\,$(\ref{eq:general_mass_matrix}) are
small.  Then, an approximate analytical solution can be found
following a two-step diagonalization procedure given in
Ref.~\cite{Choi:2006fz}. In the first step the $4\times 4$ MSSM
submatrix ${\cal M}_4$ and the new $2\times 2$ singlino--U(1)$_X$
gaugino submatrix ${\cal M}_2$ are separately diagonalised. In the
second step a block--diagonalization removes the non--zero
off--diagonal blocks while leaving the diagonal blocks approximately
diagonal up to second order, due to the weak coupling of the two
subsystems.

\subsection{The sfermion sector}

As explained in the Introduction, we assume the exotic squarks to be
substantially heavier than the MSSM fields. However the structure of
the MSSM squarks gets modified by the presence of extra U(1)$_X$. Both
the squarks and sleptons are important to our analysis and so we
briefly describe the new ingredients in the sfermion mass matrix
(neglecting the possibility of flavor and CP violation)
\begin{eqnarray}
M_{\tilde{f}}^{2} & = & \left(\begin{array}{cc}
m_{\tilde{F}}^{2}+m_{{f}}^{2}+\Delta_{\tilde{f}} &
m_{f}(A_{f}-\mu(\tan\beta)^{-2I^3_f})\\
m_{f}(A_{f}-\mu(\tan\beta)^{-2I^3_f}) &
m_{\tilde{f}}^{2}+m_{f}^{2}+\Delta_{\tilde{f}^*}\end{array}\right).
\end{eqnarray}
where $m_{\tilde{F}}$, $m_{\tilde{f}}$ are the sfermion
soft-supersymmetry breaking parameters for the quark and lepton
doublets $F=Q,L$ and singlets $f=U^c,D^c,E^c$ and $A_f$ is the
trilinear coupling, while $m_f$ is the corresponding fermion mass and
the $D$-terms receive additional U(1)$_X$ terms
\begin{equation}
  \Delta_{\tilde{f}}=M_{Z}^{2}\cos2\beta(I^{3}_f-e_{f}s_{W}^{2})+\nicefrac{1}{2}
  g_{1}^{'2}Q_{\tilde{f}}\left[v^{2}\left(Q_{1}\cos^{2}\beta+Q_{2}\sin^{2}\beta\right)
    +Q_{S}v_{S}^{2}\right]\label{delta_q}
\end{equation}
where $I^3_f$ and $e_f$ are the weak isospin and electric charge and
the U(1)$_X$ charges $Q_{\tilde{f}}$ are for the left fields.
Explicitly, we have for squarks
\begin{eqnarray}
\Delta_{\tilde{u}} & = &
M_{Z}^{2}\cos2\beta\,(\nicefrac{1}{2}-\nicefrac{2}{3}s_{W}^{2})+
\nicefrac{1}{80}g_{1}^{'2}[-v^{2}(2\sin^{2}\beta+3\cos^{2}\beta)+
5v_{S}^{2}]\nonumber \\ \Delta_{\tilde{u}^*} & = &
M_{Z}^{2}\cos2\beta\, \nicefrac{2}{3}s_{W}^{2}
+\nicefrac{1}{80}g_{1}^{'2}[-v^{2}
(2\sin^{2}\beta+3\cos^{2}\beta)+5v_{S}^{2}]\nonumber \\
\Delta_{\tilde{d}} & = &
M_{Z}^{2}\cos2\beta\,(-\nicefrac{1}{2}+\nicefrac{1}{3}s_{W}^{2})+
\nicefrac{1}{80}g_{1}^{'2}[-v^{2}(2\sin^{2}\beta+3\cos^{2}\beta)+
5v_{S}^{2}]\nonumber \\ \Delta_{\tilde{d}^*} & = &
M_{Z}^{2}\cos2\beta\,(-\nicefrac{1}{3}s_{W}^{2})+
\nicefrac{2}{80}g_{1}^{'2} [-v^{2}(2\sin^{2}\beta+3\cos^{2}\beta)+
5v_{S}^{2}]\label{delta_ud}
\end{eqnarray}
and for sleptons
\begin{eqnarray}
\Delta_{\tilde{\nu}} & = &
M_{Z}^{2}\cos2\beta\,(\nicefrac{1}{2})+
\nicefrac{2}{80}g_{1}^{'2}[-v^{2}(2\sin^{2}\beta+3\cos^{2}\beta)+
5v_{S}^{2}]\nonumber \\
\Delta_{\tilde{e}} & = &
M_{Z}^{2}\cos2\beta\,(-\nicefrac{1}{2}+s_{W}^{2})+
\nicefrac{2}{80}g_{1}^{'2}[-v^{2}(2\sin^{2}\beta+3\cos^{2}\beta)+
5v_{S}^{2}]\nonumber \\
\Delta_{\tilde{e}^*} & = &
M_{Z}^{2}\cos2\beta\,(-s_{W}^{2})+
\nicefrac{1}{80}g_{1}^{'2} [-v^{2}(2\sin^{2}\beta+3\cos^{2}\beta)+
5v_{S}^{2}]\label{delta_enu}
\end{eqnarray}
Note that here $g_1'$ is the GUT normalized $U(1)_X$ gauge coupling
analogous to the GUT normalized hypercharge gauge coupling $g_1$ in
the MSSM.

The diagonal form of the sfermion mass matrix is obtained, as usual,
by a 2x2 rotation in the $LR$ plane
\begin{equation}
M_{\tilde{f}}^{2\,
diag}=U_{\tilde{f}}^{T}M_{\tilde{f}}^{2}U_{\tilde{f}}\label{diagM_q}\end{equation}
and the mass eigenstates are defined according to
\begin{equation}
\left(\begin{array}{c} \tilde{f}_{1}\\
\tilde{f}_{2}\end{array}\right)=U_{\tilde{f}}^{\dagger}\left(\begin{array}{c}
\tilde{f}_{L}\\
\tilde{f}_{R}\end{array}\right)\label{diagU_f}\end{equation} with the
convention that $m_{\tilde{f}_{1}}\leq m_{\tilde{f}_{2}}$.

\section{Calculating the relic density}

The calculation of the neutralino LSP relic density in the MSSM is
well known \cite{EHNOS,hep-ph/9506380} and has been widely studied in
the general MSSM \cite{all} and the constrained MSSM
\cite{Kane:1993td}. The calculation of the relic density in the NMSSM
has also been extensively studied \cite{nmssmDM}. The differences
between the MSSM relic density calculation and the USSM calculation
arise through the extension of the particle spectrum and through the
new interactions that are introduced. We have implemented all new
interactions into the {\tt micrOMEGAs}\cite{micromegas} code using
{\tt LanHep}\cite{lanhep} to generate the feynman rules. {\tt
MicrOMEGAs} takes full account of all annihilation and coannihilation
processes and calculates their effect whenever they are
relevant. Nevetheless, from the form of these alterations we would
like to make some general observations before we go on to consider the
details of the calculations.

The USSM extends the neutralino sector by adding two new states to the
spectrum: the bino' and singlino components. This results in two extra
neutralinos. However for the relic density calculation we are only
interested in the lightest neutralinos, so the primary effect will be
through the magnitude of the singlino and bino' components in the
lightest neutralino. In what follows we will be interested in the
scenarios in which the lightest neutralino has a significant singlino
component and a small but non-zero bino' component. Therefore it is
informative to consider the general form of the interactions that
arise from the singlino and bino' components of the lightest
neutralino before considering specific diagrams.

The bino' component is always subdominant to the singlino component
due to the see-saw structure of the extra $2\times 2$ $\SB$ sector of
the neutralino mass matrix in Eq.~\ref{eq:general_mass_matrix}. The
form of the interactions that arise from the inclusion of the bino'
component closely mirror those of the bino component, except for the
different coupling constant and charges under the new $U(1)_X$.

The singlino component is another matter. It gives rise to a new type
of neutralino interaction from the $\lambda \hat{S}\,
\hat{H}_u\hat{H}_d$ term in the superpotential that will be seen to
dominate the annihilation processes of neutralinos with significant
singlino components. This term means that if the lightest neutralino
has significant singlino and higgsino components then it will couple
strongly to Higgs bosons with a significant $H_u$ or $H_d$ component,
usually the lighter Higgs bosons, $H_{1,2}\text{ and }A$ in the
spectrum.  Moreover, the absence of the singlet cubic term $\tilde
S^3$, in contrast the the NMSSM, implies that the singlino-dominated
LSP {\it needs} an admixture of MSSM higgsinos to annihilate to
Higgs bosons.

On the other hand, the singlino component does not interact with the
$SU(2)$ or $U(1)_Y$ gauginos. Therefore a significant singlino
component in the lightest neutralino will suppress annihilations to
$W$ or $Z_1$ bosons.

Finally, there is no coupling of the singlino component to
fermions. Thus a significant singlino component in the lightest
neutralino will also suppress annihilation to fermions.

Having noted these general features we will now consider the specific
behavior of the different annihilation diagrams.

\subsection{t-channel diagrams}

\begin{itemize}
\item Gauge boson final states
\end{itemize}

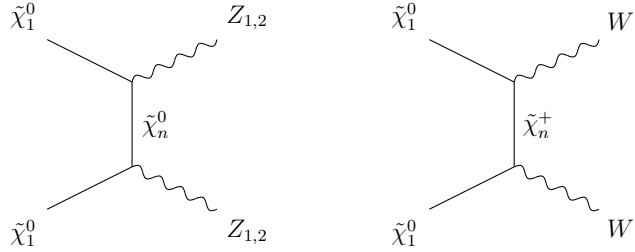
\begin{figure}[h]
\begin{center}
\scalebox{0.8}{
  \begin{picture}(360,120)
    \Text(35,105)[br]{$\neut$}
    \Line(40,100)(80,80)
    \Line(80,80)(80,40)
    \Text(85,60)[l]{$\tilde{\chi}^0_n$}
    \Line(80,40)(40,20)
    \Text(35,15)[tr]{$\neut$}
    \Photon(80,80)(120,100){2}{4}
    \Text(125,105)[bl]{$Z_{1,2}$}
    \Photon(80,40)(120,20){2}{4}
    \Text(125,15)[tl]{$Z_{1,2}$}
    \Text(215,105)[br]{$\neut$}
    \Line(220,100)(260,80) \Line(260,80)(260,40)
    \Text(265,60)[l]{$\tilde{\chi}^+_n$} \Line(260,40)(220,20)
    \Text(215,15)[tr]{$\neut$} \Photon(260,80)(300,100){2}{4}
    \Text(305,105)[bl]{$W$} \Photon(260,40)(300,20){2}{4}
    \Text(305,15)[tl]{$W$}
  \end{picture}}
\end{center}
\caption{\emph{The t-channel annihilation processes for a neutralino
    to final states containing gauge bosons.}}
\label{fig:SBtoGauge}
\end{figure}

Fig.~\ref{fig:SBtoGauge} shows the t-channel diagrams available for
annihilation of neutralinos to gauge bosons. The
$\neut\tilde{\chi}^0_j Z_i$ vertex is given in
Eq.~(\ref{eq:NeutNeutZ}). Note that the coupling of neutralinos to the
$Z$ component of the $Z_1$ state is precisely that of the MSSM $\neut
\tilde{\chi}^0_j Z$ coupling. As the $Z$ component dominates the $Z_1$
state, a singlino dominated LSP will not annihilate strongly to $Z_1$
bosons.

In contrast there is a strong coupling from the MSSM-higgsino
components as well as the singlino component to the $Z'$ component of
the $Z_i$ state. Notice also that the MSSM-higgsino components of the
LSP enter with the same sign in the coupling to the $Z'$, unlike in
the coupling to the $Z$, where they tend to cancel each other. As the
$Z_2$ boson is dominantly $Z'$ any LSP with a non-zero higgsino or
singlino fraction will annihilate to $Z_2$ bosons when such a final
state is kinematically allowed. Unfortunately the $Z_2$ is required to
be heavy by experimental limits, so annihilation of the lightest
neutralinos to final states involving one $Z_2$ is hard to achieve and
annihilation to two $Z_2$ bosons is impossible.

The second diagram of Fig.~\ref{fig:SBtoGauge} shows the t-channel
annihilation to $W^\pm$ final states. Eqs.~(\ref{eq:WChargNeuta}) and
(\ref{eq:WChargNeutb}) give the relevant coupling and show that the
singlino and bino' components do not couple to the wino component of
charginos or to the $W^\pm$ bosons. This means that a large singlino
or bino' component in the LSP will suppress annihilation to $W^\pm$
bosons in the final state.

\begin{itemize}
\item Higgs boson final states
\end{itemize}

\begin{figure}[h]
\begin{center}
\scalebox{0.8}{\begin{picture}(540,120)
    \Text(35,105)[br]{$\neut$}
    \Line(40,100)(80,80)
    \Line(80,80)(80,40)
    \Text(85,60)[l]{$\tilde{\chi}^0_n$}
    \Line(80,40)(40,20)
    \Text(35,15)[tr]{$\neut$}
    \DashLine(80,80)(120,100){5}
    \Text(125,105)[bl]{$H_i$}
    \DashLine(80,40)(120,20){5}
    \Text(125,15)[tl]{$H_j$}
    \Text(215,105)[br]{$\neut$}
    \Line(220,100)(260,80)
    \Line(260,80)(260,40)
    \Text(265,60)[l]{$\tilde{\chi}^0_n$}
    \Line(260,40)(220,20)
    \Text(215,15)[tr]{$\neut$}
    \DashLine(260,80)(300,100){5}
    \Text(305,105)[bl]{$A$}
    \DashLine(260,40)(300,20){5}
    \Text(305,15)[tl]{$A$}
    \Text(395,105)[br]{$\neut$}
    \Line(400,100)(440,80)
    \Line(440,80)(440,40)
    \Text(445,60)[l]{$\tilde{\chi}^+_n$}
    \Line(440,40)(400,20)
    \Text(395,15)[tr]{$\neut$}
    \DashLine(440,80)(480,100){5}
    \Text(485,105)[bl]{$H^+$}
    \DashLine(440,40)(480,20){5}
    \Text(485,15)[tl]{$H^-$}
  \end{picture}}
\end{center}
\caption{\emph{The t-channel annihilation processes for a neutralino
    to final states involving scalar Higgs bosons, pseudoscalar Higgs
    bosons or charged Higgs bosons respectively.}}
\label{fig:SBtoHiggs}
\end{figure}
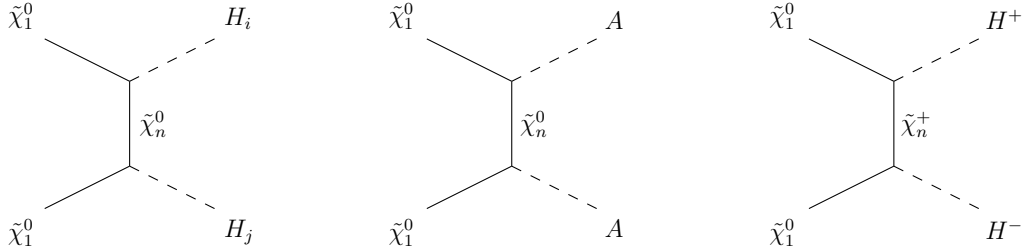

Fig.~\ref{fig:SBtoHiggs} shows the available t-channel processes for
the annihilation of neutralinos to final state Higgs bosons. Due to
the $\lambda \hat{S}\, \hat{H}_u\hat{H}_d$ term in the superpotential
and the D-terms there are significant differences between these
diagrams in the USSM and the MSSM. The $\neut\tilde{\chi}^0_i H_j$
vertex given in Eq.~(\ref{hxx}) is the relevant vertex in this first
diagram.

First note that the bino' component of one neutralino couples with the
higgsino component of the other and the $H_{u,d}$ component of the
final state Higgs boson in the same way as the equivalent coupling of
the bino or wino components. In addition there is an extra term which
couples the bino' component of one neutralino to the singlino
component of the other and to the singlet component of the Higgs boson
in the final state. This means that if the lightest neutralino is
dominantly singlino, it will annihilate to final state Higgs bosons
with a significant singlet component through the exchange of a
neutralino with a significant bino' component in the
t-channel. Unfortunately these processes are disfavored for the same
reason as annihilation to final states containing a $Z_2$. The Higgs
boson with a significant singlet component will have a mass comparable
to the $Z_2$ boson and thus a final state with two such Higgs bosons
will be impossible and even one will often be kinematically ruled out.

Of more interest is the term in this vertex that couples a singlino
component of one neutralino to a higgsino component of the other
neutralino and the $H_{u,d}$ components of the Higgs boson with a
strength $\lambda$. If the lightest neutralino is dominantly singlino
then two LSPs can exchange a dominantly higgsino neutralino in the
t-channel to produce two Higgs bosons in the final state. This is a
channel that is always present if the lightest neutralinos are heavy
enough to produce two light Higgs bosons in the final
state. Obviously, if both $H_1$ and $H_2$ are lighter than the
lightest neutralino then there will be more available channels. As the
singlino couples predominantly to Higgs states, this channel provides
the strongest annihilation mechanism for a neutralino with a large
singlino component. This amplitude will be maximised for three
degenerate mixed state neutralinos with strong higgsino and singlino
components that are heavier in mass than the lightest two Higgs
states. The addition of this vertex also allows for a new annihilation
process for a dominantly higgsino neutralino through the exchange of a
t-channel neutralino with a substantial singlino component.

The middle diagram of Fig.~\ref{fig:SBtoHiggs} shows the annihilation
to final state pseudoscalar Higgs bosons. The relevant vertex is given
in Eq.~(\ref{Axx}). The first line gives the familiar MSSM vertex for
the coupling of a $\tilde{B}$ or $\tilde{W}$ component of a neutralino
to a higgsino component and a pseudoscalar Higgs. This is modified by
an overall factor of $\cos\phi$ which determines the magnitude of the
MSSM-like components of the pseudoscalar Higgs over the singlet
contribution. As $\sin\phi\approx v\cos\beta/v_S$, the suppression
from $\cos\phi$ is small. This is the same as saying that the
pseudoscalar Higgs generally only has a very small singlet
component. The analogue of the $\tilde{W},~\tilde{B}$ interaction
terms appears for the bino'. The bino' component also couples to the
singlino component of the second neutralino and the singlet component
of the final state pseudoscalar Higgs. This term is $\sin\phi$
suppressed due to the small singlet component of the $A$ bosons in the
final state. These interactions determine the strength of the
annihilation of a dominantly gaugino LSP to pseudoscalar Higgs bosons
through the exchange of a dominantly higgsino (or singlino)
neutralino.

More interesting contributions come from the $\lambda \hat{S}\,
\hat{H}_u\hat{H}_d$ term in the superpotential. These provide a
$A\tilde{H}_u\tilde{H}_d$ coupling, albeit suppressed by a factor of
$\sin\phi$. Such a coupling does not appear in the MSSM. There is also
a term that couples $A\tilde{S}\tilde{H}_{u,d}$ with no $\sin\phi$
suppression. Once again this produces a strong annihilation channel
for a neutralino with a substantial singlino component through
t-channel neutralino exchange where the neutralino exchanged in the
t-channel must have a significant higgsino component. This is the
analogue of the process we discussed in some detail for the scalar
Higgs final states and will, kinematics allowing, give a strong
annihilation channel for a dominantly singlino neutralino as long as
there is a light neutralino in the spectrum with a substantial
higgsino component to be exchanged in the t-channel.

The final diagram of Fig.~\ref{fig:SBtoHiggs} shows annihilation to
charged Higgs boson final states.  The relevant vertex is given in
Eq.~(\ref{xx+H-}). The vertex includes a $\tilde{B}'$ interaction that
parallels the familiar $\tilde{B}$ and $\tilde{W}$ interactions to the
higgsino component of the chargino and a charged Higgs boson. There is
also a term that arises from the $\lambda \hat{S}\,
\hat{H}_u\hat{H}_d$ superpotential term. This allows for a neutralino
with a substantial singlino component to annihilate to charged Higgs
bosons via t-channel chargino exchange as long as there are light
charginos with a significant higgsino component and the final state
charged Higgs bosons are kinematically allowed. In contrast to the
previous two diagrams, this one does not add an extra annihilation
channel for a dominantly higgsino neutralino. In the first two
diagrams there is the new possibility in which a dominantly singlino
neutralino is exchanged in the t-channel. In the third diagram there
is no such process as there is no singlino component in the charginos.

From an analysis of the processes with Higgs bosons in the final state
we see that there will be a strong annihilation cross-section for a
neutralino with a large singlino component to light Higgs bosons if
there is a light neutralino with a substantial higgsino component in
the spectrum and the Higgs boson final states are kinematically
allowed. We also note that the $\lambda \hat{S}\, \hat{H}_u\hat{H}_d$
allows for new couplings between neutralinos and Higgs bosons that
will alter the annihilation of dominantly higgsino neutralinos with
respect to their behavior in the MSSM.

\begin{itemize}
\item Mixed boson final states
\end{itemize}

It is quite possible to have an unmatched pair of bosons in the final
state of a t-channel annihilation diagram. We do not need to go
through the details of all possible diagrams. Instead we just note
that a neutralino with a significant singlino component will
dominantly annihilate to final states made up of Higgs bosons. The
strength of such channels will depend upon the size of the singlino
component in the lightest neutralino, the mass of the neutralinos with
substantial higgsino components that will be exchanged in the
t-channel, and the mass of the final state Higgs bosons.

\begin{itemize}
\item Fermion final states
\end{itemize}

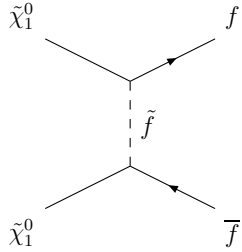
\begin{figure}
\begin{center}
\scalebox{0.8}{\begin{picture}(180,120)
    \Line(40,100)(80,80)
    \Text(35,105)[br]{$\neut$}
    \DashLine(80,80)(80,40){5}
    \Text(85,60)[l]{$\tilde{f}$}
    \Line(80,40)(40,20)
    \Text(35,15)[tr]{$\neut$}
    \ArrowLine(80,80)(120,100)
    \Text(125,105)[bl]{$f$}
    \ArrowLine(120,20)(80,40)
    \Text(125,15)[tl]{$\overline{f}$}
  \end{picture}}
\end{center}
\caption{\emph{The t-channel annihilation process for neutralinos
     to fermions.}}
\label{fig:SBtoFerm}
\end{figure}

Finally we consider the t-channel annihilation diagram to final state
fermions through the diagram given in Fig.~\ref{fig:SBtoFerm}. The
squark vertices are given in Eqs.~(\ref{xSqQa}) and (\ref{xSqQb}). The
couplings of the bino and wino components of the neutralino are the
same as in the MSSM. Note that there is an extra coupling of the bino'
component of the neutralino to the squark-quark pair that is of the
same order of magnitude as for the $\tilde B$. As the bino' is only
ever a subdominant component of the lightest neutralino, and as the
annihilation to fermions is relatively weak in the first place, we can
expect that interactions of this form will have little impact on the
annihilation cross-section. However, if the lightest neutralino is too
light to annihilate to final state Higgs bosons, this channel will
remain open and can dominate though it will give a relic density well
in excess of that measured by WMAP.

\subsection{s-channel diagrams}

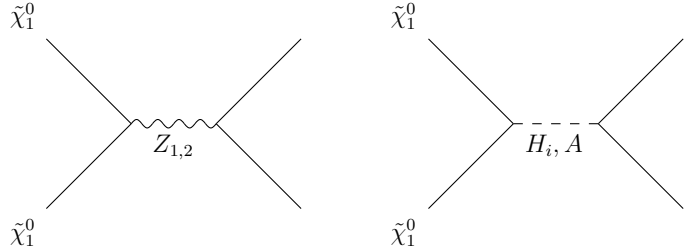
\begin{figure}
  \begin{center}
    \scalebox{0.8}{\begin{picture}(380,120)
    \Line(20,100)(60,60)
    \Text(15,105)[br]{$\neut$}
    \Line(60,60)(20,20)
    \Text(15,15)[tr]{$\neut$}
    \Photon(60,60)(100,60){2}{4}
    \Text(80,55)[t]{$Z_{1,2}$}
    \Line(100,60)(140,100)
    \Line(140,20)(100,60)
    \Line(200,100)(240,60)
    \Text(195,105)[br]{$\neut$}
    \Line(240,60)(200,20)
    \Text(195,15)[tr]{$\neut$}
    \DashLine(240,60)(280,60){5}
    \Text(260,55)[t]{$H_i,A$}
    \Line(280,60)(320,100)
    \Line(280,60)(320,20)
    \end{picture}}
  \end{center}
  \caption{\emph{The annihilation processes for a neutralino through
      s-channel Higgs and $Z_i$ bosons, where we do not specify the
      precise particles in the final state.}}
  \label{fig:schan}
\end{figure}

Fig.~\ref{fig:schan} shows the possible s-channel processes available
for the annihilation of a pair of neutralinos. The first diagram
shows the annihilation through and intermediate $Z_i$ gauge
boson. The relevant coupling of two neutralinos to a $Z_i$ is given in
Eq.~(\ref{eq:NeutNeutZ}). As before we note that the singlino
component of the neutralino only couples to the $Z'$ component of the
$Z_i$ gauge boson. This means that if the lightest neutralino has a
significant singlino component, then annihilations through an
s-channel $Z_1$ will be suppressed as the $Z'$ component of the $Z_1$
is required to be very small. On the other hand, the $Z_2$ has a large
$Z'$ component. Therefore a lightest neutralino with a substantial
singlino component will annihilate through an s-channel $Z_2$.

The second diagram shows the annihilation of neutralinos through an
s-channel scalar or pseudoscalar Higgs boson. The relevant couplings
are given in Eq.~(\ref{hxx}) and Eq.~(\ref{Axx})
respectively. Consider first the case in which the s-channel Higgs
boson is dominantly composed of the singlet Higgs. Since the bino or
wino components of the lightest neutralino only couple to the
non-singlet Higgs components of the s-channel Higgs a light neutralino
that is dominantly wino or bino will not annihilate strongly through a
dominantly singlet Higgs. However there is a coupling of the singlet
component of the Higgs boson to the higgsino components of the
lightest neutralino. This provides a strong channel when on-resonance
for annihilation of a light neutralino with a large higgsino
component. There is also a strong coupling if the lightest neutralino
has significant bino' and singlino components. Thus we expect a light
neutralino with strong mixing between higgsino, singlino and bino'
terms to annihilate strongly through s-channel heavy Higgs exchange
where the heavy higgs has a large singlet Higgs component.

If the s-channel Higgs boson does not have a large singlet Higgs
component then the story is somewhat different. In this case the light
neutralino needs to have a significant higgsino fraction along with a
substantial contribution from one of the other non-higgsino
states. This situation is mirrored in the case of the pseudoscalar
Higgs.

From this we see that we have a new annihilation channel for
neutralinos with a significant higgsino fraction through a dominantly
singlet Higgs in the s-channel. We also see that a light neutralino
with a substantial singlino-higgsino mixture will annihilate strongly
through the whole range of s-channel Higgs exchange processes.

\subsection{Coannihilation}

As well as the annihilation of two identical neutralinos, it is often
the case that coannhilation between the LSP and the NLSP (and
sometimes even heavier states) can be important. This process is
normally important for a dominantly MSSM-higgsino or wino neutralino
LSP. In these situations there is an automatic near degeneracy in the
mass of the lightest neutralino with the mass of the lightest chargino
and, in the case of the higgsinos, also with the next-to-lightest
neutralino. A dominantly singlino LSP does not have an automatic
degeneracy with other states. However, it is possible for a singlino
neutralino to be exactly degenerate with other states - something that
does not happen in the MSSM due to the signs of the terms in the
neutralino mixing matrix. In these cases we would expect the effect of
coannihilation to be important.

Therefore we expect coannihilation processes to only be significant in
regions of the parameter space where we move from one type of LSP to
another as this indicates a degeneracy in the mass of the LSP and
NLSP. We also expect to see the standard large coannihilation
contributions for a predominantly MSSM-like higgsino LSP or
predominantly wino LSP.

\section{Elastic scattering of neutralinos from nuclei}

The direct cold dark matter search experiments, such as
DAMA/LIBRA, CDMS, ZEPLIN, EDELWEISS, CRESST, XENON, WARP
\cite{directDM}, aim at detecting dark matter particles through
their elastic scattering with nuclei. This is complementary to
indirect detection efforts, such as GLAST, EGRET, H.E.S.S.
\cite{indirectDM},  which attempt to observe the annihilation
products of dark matter particles trapped in celestial bodies.

\begin{figure}[h]
\begin{center}\includegraphics[scale=0.6]{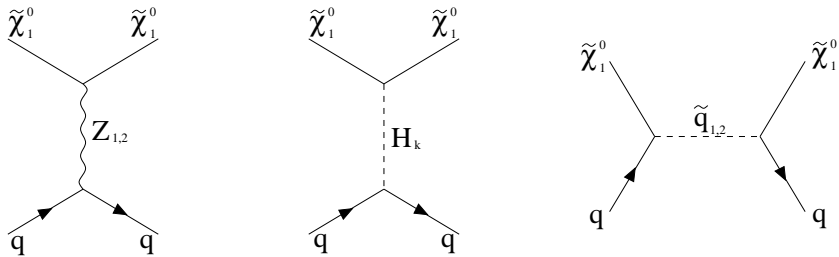}\end{center}
  \caption{\emph{Diagrams contributing to the
lightest neutralino scattering from a quark.}\label{fig:lsp-n_diag}}
\end{figure}

Since we assume the LSP to be the lightest neutralino
$\widetilde{\chi}_{1}^{0}$, we consider the elastic scattering of the
lightest neutralino from nuclei.  The elastic scattering is mediated
by the $t$-channel $Z_i$ and Higgs $H_{k}$ exchange, as well as the
$s$-channel squark $\tilde{q}_j$ exchange, as depicted in
Fig.\ref{fig:lsp-n_diag} for $\tilde{\chi}^0_1\,q$ scattering.  There
are also important contributions from interactions of neutralinos with
gluons at one loop~\cite{Griest:1988ma,Drees:1993bu}.

The extended particle content and new couplings present in the USSM
model have also a direct effect on the elastic cross section
calculations, as discussed in the previous chapter.

The elastic cross section for neutralino scattering from a nucleus can be broken into
a spin-independent (SI) and a spin-dependent (SD) part,
\begin{equation}
\sigma=\sigma_{_{\textrm{SI}}}+\sigma_{_{\textrm{SD}}}\, ,
\end{equation}
each of which can be expressed in terms of the elastic scattering of
neutralino from individual nucleons in the nuclei. In the limit of
zero-momentum transfer they can be written as~\cite{Jungman:1995df}
\begin{eqnarray}
&&\sigma_{_{\textrm{SI}}}=\frac{4\, m_{r}^{2}}{\pi}\,\left[\,
Zf_{p}+(A-Z)\, f_{n}\,\right]^{2},\\
&&\sigma_{\textrm{SD}}=\frac{32 m_{r}^{2}}{\pi}\, G_{F}^{2}\, J(J+1)\,\,
\Lambda^{2},
\end{eqnarray}
where $Z$ and $A$ are atomic number and mass of the nucleus, $J$ is
the total nucleus angular momentum and $m_{r}$ is the reduced
neutralino-nucleus mass. Note that the spin-independent part benefits
from coherent effect of the scalar couplings, which leads to cross
section and rates proportional to the square of the atomic mass of the
target nuclei.

The spin-dependent quantity $\Lambda$ is given by
\begin{eqnarray} \Lambda_{n}=\frac{1}{J}\left[\,
\langle S_{p}\rangle \, \sum_{q=u,d,s}{\frac{A_{q}}{\sqrt{2}\, G_{F}}}\,\Delta_{q}^{p}
+
\langle S_{n}\rangle\, \sum_{q=u,d,s}{\frac{A_{q}}{\sqrt{2}\, G_{F}}}\,\Delta_{q}^{n}
\right]
\end{eqnarray}
where $\langle S_{p}\rangle$ and $\langle S_{n}\rangle$ are the
expectation values of the spin content of the proton and neutron group
in the nucleus, while $\Delta_{q}^{p}$ and $\Delta_{q}^{n}$ are the
quark spin content of the proton and neutron, respectively.

For the spin-independent part, the effective couplings of the LSP
neutralino to proton and neutron $f_{p}$ and $f_{n}$ are more
complicated. In the limit of $m_{\tilde{\chi}_{1}^{0}}\ll
m_{\tilde{q}}$ and $m_{q}\ll m_{\tilde{q}}$, which we will later
confine to, they simplify and can be approximated as:\begin{eqnarray}
\frac{f_{p,n}}{m_{p,n}}=\sum_{q=u,d,s}f_{_{{\textrm{T}}_{q}}}^{p,n}\,\frac{B_{q}}{m_{q}}+\frac{2}{27}\,
f_{_{{\textrm{T}}_{G}}}^{p,n}\sum_{q=c,b,t}\frac{B_{q}}{m_{q}}
\label{eq:chi2pn}
\end{eqnarray}
The first term in Eq.(\ref{eq:chi2pn}) corresponds to interactions
with the quarks in the target nuclei, while the second term
corresponds to interactions with the gluons in the target through a
quark/squark loop diagram, and
\begin{eqnarray}
f_{_{{\textrm{T}}_{G}}}^{p,n}=1-\sum_{q=u,d,s}f_{{\textrm{T}}_{q}}^{p,n}.
\end{eqnarray}
Finally, the effective Lagrangian for elastic scattering of
neutralinos from quarks in the non-relativistic limit can be written
as a sum of axial-vector (spin-dependent) and scalar
(spin-independent) terms:
\begin{eqnarray}
{\mathcal{L}}_{\textrm{eff}}=A_{q}\,(\bar{\chi}_{1}\gamma^{\mu}\gamma_{5}\chi_{1})\,
(\bar{q}\gamma_{\mu}\gamma_{5}q)+B_{q}\,(\bar{\chi}_{1}\chi_{1})\,(\bar{q}q)
\end{eqnarray}
The effective couplings $A_{q}$ and $B_{q}$ are given by:
\begin{eqnarray} \label{eq:coefficients}
&& A_q = \frac{g_2^2}{16}
\sum_{i=1,2} \frac{\left|\, B_q^{i L}\right|^2 + \left|\,
B_q^{iR}\right|^2 } { m_{\tilde{q_i}}^2 - (m_{\tilde{\chi_0}} -m_q )^2
} - \frac{G_F}{\sqrt{2}} \left[\left|N_{13}\right|^2
-\left|N_{14}\right|^2\right] I^3_{q}\nonumber\\
&& ~~~~~~~
-\frac{g'^2_1}{4 m^2_{Z'}}\, \left[\,{Q}_1 |N_{13} |^{\,2} +
{Q}_2|N_{14} |^{\,2} +{Q}_s|N_{15} |^{\,2}\right]\,({Q}_{Q} +
{Q}_{\bar{q}})\, \label{eq:axial}\\[2mm]
&& B_q = -\frac{g_2^2}{8} \sum_{i=1,2} \frac{\re (B_q^{iL} {B_q^{iR}}^*)
} { m_{\tilde{q_i}}^2 - (m_{\tilde{\chi_0}}-m_q )^2 } \nonumber\\ &&
~~~~~~~ - \frac{h_q }{2\sqrt{2}} \sum_{k=1}^{3}
\frac{\re(G_k)+\re(G'_k)+\re(G''_k)}{m_{H_k}^2}\;
\left\{\begin{array}{ll}
 {\cal O}'_{1k} &\textrm{for $q=d,s,b$  }\\
 {\cal O}'_{2k} &\textrm{for $q=u,c,t$ } \end{array} \right .
\label{eq:scalar}
\end{eqnarray}
In this expressions  we have neglected a small $Z$-$Z'$ mixing.

The first terms in both effective couplings come from squark exchange
diagrams. The neutralino-squark-quark couplings $B^{iL}_q,\, B^{iR}_q$
are given in Appendix~B. As seen in
Eqs.~(\ref{eq:binoup},\ref{eq:binodown}), they receive a contribution
form the bino' component $N_{16}$.

The second and the third terms in (\ref{eq:axial}) come from the $Z$
and $Z'$ exchanges, respectively, where the latter contains a term due
to the singlino component, $N_{15}$.  The second term in the form
factor $B_q$ receives contributions from three scalar Higgs boson
exchanges. Each contains an MSSM-like term, $G_k$, as well as the new
terms $G'_k$ and $G''_k$, ($k=1,2,3$)
 \begin{eqnarray}
G_{k} & = & g_2(N_{12}-t_{W}\, N_{11})(N_{14}\, {\cal O}'_{2k} - N_{13}\,
{\cal O}'_{1k})\,\,\,\,\,\nonumber \\
G'_{k} & = &
     -2\, g'_1\, N_{16}\,({Q}_{1}\,
       N_{13}\, {\cal O}'_{1k} +{Q}_{2}\, N_{14}\, {\cal O}'_{2k}\,+
       {Q}_{S}\, N_{15}\,{\cal O}'_{3k}) \nonumber \\
G''_{k} & = & \sqrt{2}\lambda\, \left[
        N_{15}\,( N_{13}\, {\cal O}'_{2k}+ N_{14}\,{\cal O}'_{1k})
        + N_{13}N_{14} {\cal O}'_{3k}\,\right]\hskip1cm\label{eq:Gterms}
\end{eqnarray}

The $G'_k$ piece is generated by the $g'_1\tilde B'(\tilde H_i H_i+
\tilde S S)$ couplings from the extra U(1)$_X$ $D$-terms, while the
$G''_k$ is induced by the $\lambda\tilde H_i (\tilde S H_j + \tilde
H_j S)$ couplings (here we follow the conventions and notations of
Ref.~\cite{Choi:2000kh}, properly extended to the USSM model
\cite{DJ}).

\section{Results}
Now that we have introduced the model we move on to study the details
of the dark matter phenomenology within the USSM parameter space.

\subsection{Defining a parameter range}

Before we study the phenomenology we need to define the parameter
range we are interested in. The USSM extends the number of free
parameters over those in the MSSM by the set:
\begin{equation*}
M_1',~g_1',~\lambda,~A_{\lambda},~v_S.
\end{equation*}
These parameters are constrained by a number of different factors.

We fix $g_1'=g_1$ as we wish to maintain gauge coupling unification
and the two $U(1)$ gauge couplings run with identical RGEs.

The parameters $v_S,~\lambda,~A_\lambda$ appear in the determination
of particle masses. Therefore we determine these by setting the
corresponding masses. First of all, we wish to keep $v_S$ low to
maximise the region of parameter space in which there is a light
singlino/bino' LSP. If $M_1'=0$ then there are two degenerate
singlino/bino' neutralinos with a mass $Q'_S g'_1v_S$. However, we do
not have the freedom to set $v_S$ arbitrarily low since from
Eqs.~(\ref{eq:MzMz'Delta}) we see that low $v_S$ would require a light
$Z_2$ mass and a large $Z$-$Z'$ mixing incompatible with the LEP and
Tevatron limits. Adopting
\begin{equation}
  m_s=g'_1 v_S=1200~GeV,
\end{equation}
together with assumed $\tan\beta=5$, gives $M_{Z_2}=949$ GeV and
$\sin_{ZZ'}=3~10^{-3}$ which is consistent with current constraints.
We use this to set the magnitude of the $v_S$ in all that follows.

With $v_S$ set, our choice of $\lambda$ will set the size of $\mu$
through the relation
\begin{equation}
  \mu=\lambda\frac{v_S}{\sqrt{2}}.
\end{equation}
Note that $\lambda$ is a coupling and so cannot be too large.  An
upper limit on $\lambda < 0.7$ at a given value of $v_S$ results in a
corresponding maximum value on $\mu$, and consequently
$\mu<m_{S,Z'}$. As a result, $m_{\neut}<m_{Z',S}$ will always be
satisfied which has important implications for the available dark
matter annihilation channels. It also justifies our earlier claim that
there will always be light charginos and higgsinos in the spectrum if
the $Z'$ mass is low.

We set the size of $A_\lambda$ by setting the mass of the pseudoscalar
Higgs. From Eq.~(\ref{m_A}) we see that once the VeV of $S$ and
$\lambda$ have been set, the mass of $m_A$ only depends upon $\tanb$
and $A_\lambda$. As we are keeping $\tanb$ fixed, we can use
$A_\lambda$ to set the psuedoscalar Higgs mass.

The familiar MSSM parameters are also relevant to the details of both
the relic density calculation and the direct detection
phenomenology. The most important parameters are those that appear in
the neutralino mass matrix - $M_1$ and $M_2$. We keep the ratio
$M_1:M_2=1:2$ for simplicity, but there are as many ways to break this
relation in the USSM as in the MSSM.

Finally we must set $M_1'$. In what follows we take $M_1'$ as a free
variable and scan over a range of values. In our first study we take
$M_1'$ to be independent of the other gaugino masses, as in the study
of Ref.~\cite{Choi:2006fz} where the collider phenomenology has been
discussed. This will complement Ref.~\cite{Choi:2006fz} with the dark
matter calculations. On the other hand, it is also interesting to
consider a scenario in which soft SUSY breaking gaugino masses are
unified, namely $M_1:M'_1:M_2=1:1:2$ and we do this in our second
scenario. This allows us to organize our studies in the following way:
\begin{itemize}
\item{scenario A: } $M'_1$ arbitrary  ;
\item{scenario B: }  unified gaugino masses $M'_1=M_1=M_2/2$ .
\end{itemize}

To calculate the relic density we need to set the rest of the particle
spectrum. To do this we fix the pseudoscalar Higgs mass $m_A=500$~GeV
and for sfermion masses we take a common mass of
$m_{Q,u,d,L,e}=800$~GeV, and a common trilinear coupling $A=1$~TeV,
while the gluino mass is determined assuming unified gaugino masses at
the GUT scale. We have set the squarks and sleptons to be heavy as
this allows for a clearer analysis of the annihilation properties of
the neutralinos.

For the direct dark matter searches, there are large uncertainties in
the spin-dependent and spin-independent elastic cross section
calculations due to the poor knowledge of the quark spin content of
the nucleon and quark masses and hadronic matrix elements. These
uncertainties have recently been discussed in
Ref.~\cite{Ellis:2008hf}, from where we calculate the central values
of $f^{p,n}_{Tq}$:
\begin{eqnarray}
 &  & f_{_{{\textrm{T}}_{u}}}^{p}=0.027,\qquad f_{_{{\textrm{T}}_{d}}}^{p}=0.039,
 \qquad f_{_{{\textrm{T}}_{s}}}^{p}=0.36\nonumber \\
 &  & f_{_{{\textrm{T}}_{u}}}^{n}=0.0216,\qquad f_{_{{\textrm{T}}_{d}}}^{n}=0.049,
 \qquad f_{_{{\textrm{T}}_{s}}}^{n}=0.36
\end{eqnarray}
and $\Delta^{p,n}_q$:
\begin{eqnarray}
&  & \Delta_{u}^{p}=+0.84,\qquad\Delta_{d}^{p}=-0.43,\qquad\Delta_{s}^{p}=-0.09
\nonumber \\
&  & \Delta_{u}^{n}=-0.43,\qquad\Delta_{d}^{n}=+0.84,\qquad\Delta_{s}^{n}=-0.09
\end{eqnarray}

\subsection{Scenario A:  $M'_1$ arbitrary}
In this scenario we take $M'_1$ as an arbitrary parameter with the
MSSM gaugino parameters fixed at $\mu=300$~GeV, $M_1=M_2/2=750$~GeV.

\begin{table}[ht]
  \begin{center}
    \begin{tabular}{|l|l|}
      \hline
      Parameter & Value\\
      \hline
      $M_1$ & 750~GeV\\
      $M_2$ & 1500~GeV\\
      $\mu$ & 300~GeV\\
      $M_1'$ & 0-20 TeV\\
      \hline
      $\langle S \rangle$ & 2607.61~GeV\\
      $\lambda$ & 0.163\\
      $A_{\lambda}$ & 160~GeV\\
      \hline
    \end{tabular}
  \end{center}
  \caption{\emph The parameters taken for the neutralino sector in the
    scan with $\mu=300$~GeV, $m_A=500$~GeV,
    $\tanb=5$. \label{tab:mu,300}}
\end{table}

\subsubsection{Mass spectrum}
\label{scenA-ms}

\begin{figure}[h]
\begin{center}\includegraphics[scale=0.6]{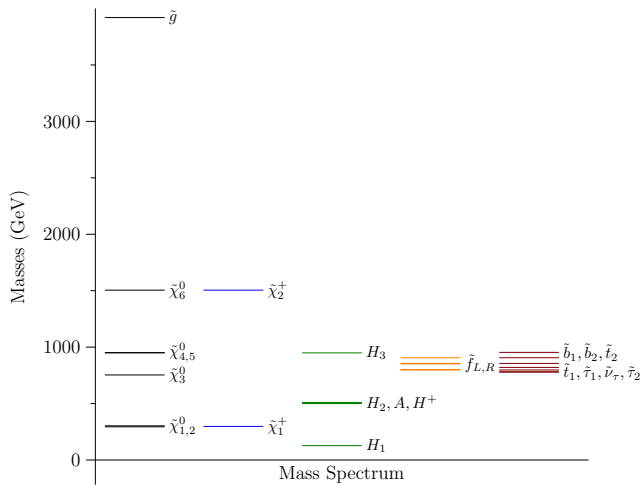}\end{center}
  \caption{\emph{The mass spectrum for
  $M_1'=0$~GeV}\label{fig:0-20MassSpec}}
\end{figure}

With these parameters we calculate the resulting mass spectrum at a
given value of $M'_1$.  The mass spectrum for $M_1'=0$~GeV is shown in
Fig.~\ref{fig:0-20MassSpec}. In the Higgs sector we have a light Higgs
at 127~GeV, a heavier scalar, pseudoscalar and charged Higgses around
500~GeV and a dominantly singlet Higgs at 949~GeV. Sfermions are
located between 750 to 950 GeV.  The chargino sector consists of a
higgsino-like chargino around $300$~GeV and a wino-like chargino at
$1500$~GeV. Since the mixing between the MSSM-like and the
bino'/singlino at $M'_1=0$ GeV is numerically small, the spectrum of
neutralinos can qualitatively be understood by separately
diagonalizing the 4x4 and 2x2 neutralino mass sub-matrices. Thus to a
good approximation we have (according to ascending (absolute) masses
for $M'_1=0$ GeV) a pair of nearly degenerate, maximally mixed MSSM
higgsinos at 300~GeV (first two states), an MSSM bino at 750~GeV (the
third), a pair of nearly degenerate, maximally mixed singlino/bino'
neutralinos at 949~GeV (the fourth and the fifth) and an MSSM wino at
1500~GeV (the sixth state).

\begin{figure}[h]
  \begin{center}
    \includegraphics[scale=0.5]{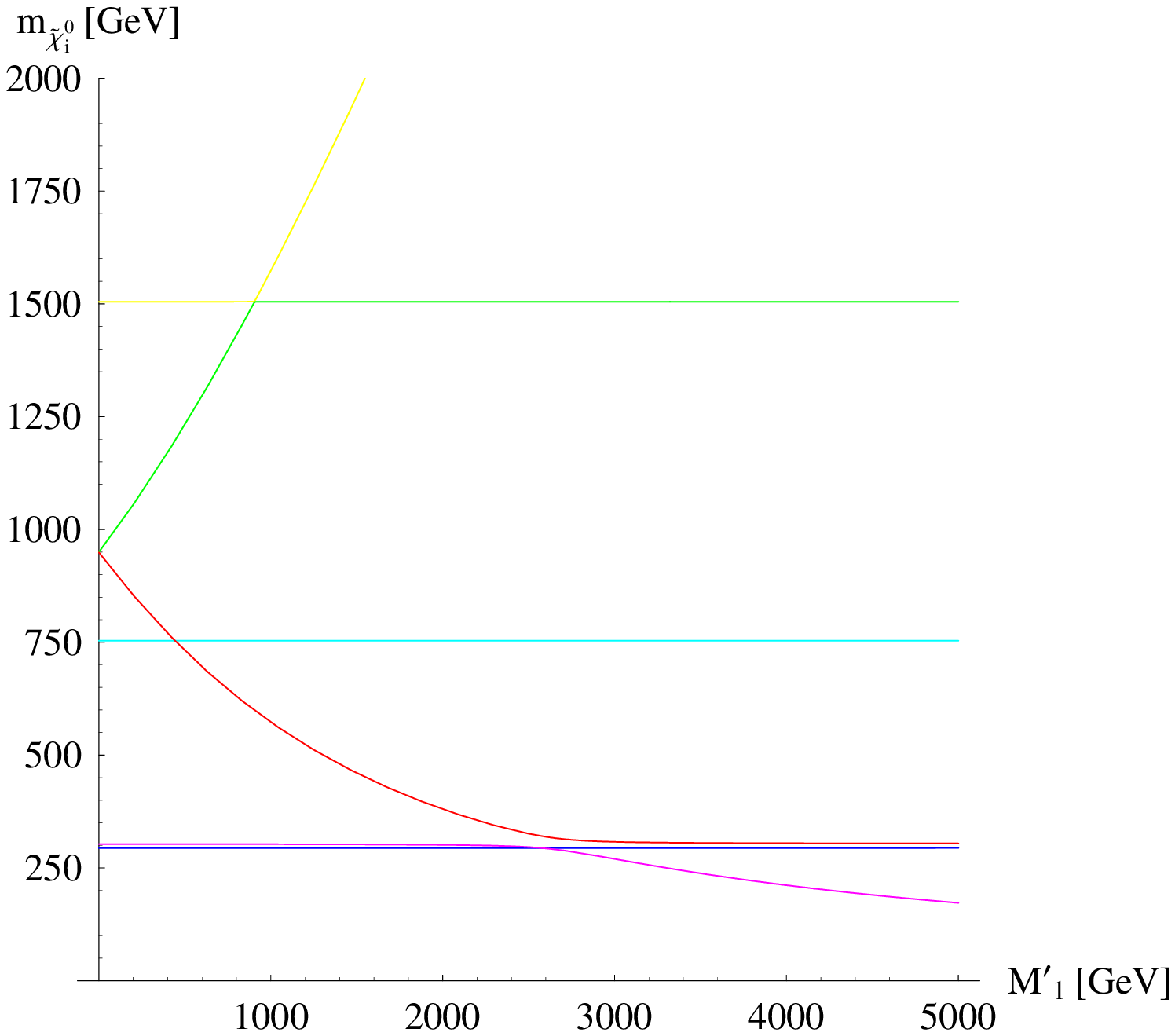}~\includegraphics[scale=0.5]{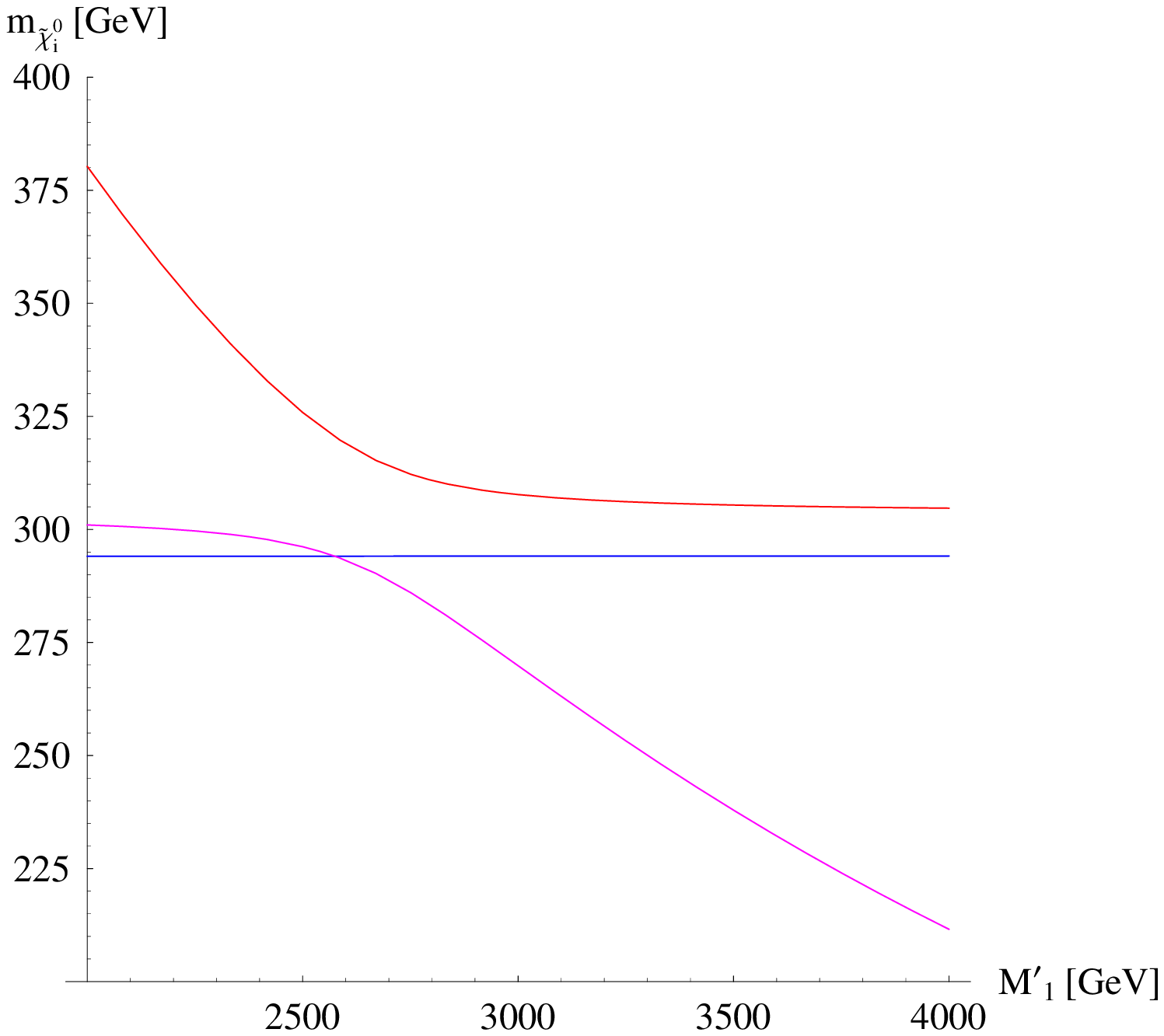}
  \end{center}
  \caption{\emph{The neutralino mass spectrum for varying $M_1'$, for
  the parameter choices in table.~\ref{tab:mu,300}. The right panel is
  a magnified part of the left one.}\label{fig:0-5NeutMasses}}
\end{figure}
To understand the change of neutralino masses and of their composition
as a function of $M'_1$ it is instructive the follow their analytic
evolution as $M'_1$ is turned on. The see-saw structure of the 2x2
singlino/bino' submatrix forces the two nearly degenerate, mixed
singlino/bino' states to move apart: the lighter one (the fourth) gets
lighter, and the mass of the other (the fifth one) heavier as $M'_1$
increases. The MSSM-like states do not evolve much, unless the mass of
one of the new states comes close to one of the MSSM, where a strong
mixing may occur. For the mixing to be important not only the
(absolute) masses must come close, but also the mass-eigenstates must
belong to eigenvalues of the same sign. It is obvious from the see-saw
structure that the heavier singlino/bino' state (the fifth one)
belongs to the positive and the lighter (the fourth) to the one
negative eigenvalue. Similarly the lighter of the two nearly
degenerate MSSM higgsinos (the first state) belongs to the positive,
and the other (the second) to negative eigenvalue.

As $M'_1$ increases the (absolute) mass of the fourth state gets
closer to the third, however they do not mix since they belong to
eigenvalues of opposite sign. In left panel of
Fig.~\ref{fig:0-5NeutMasses} the lines representing these two states
pass each other at $M'_1\sim 450$ GeV. The bino, which is the third
state according to the mass ordering below 450 GeV, becomes the fourth
one when $M'_1$ passes 450 GeV.  On the other hand when $M'_1$
approaches 900 GeV and the mass of the fifth state gets close to the
sixth one, strong mixing occurs between these states -- the two lines
representing these states in Fig.~\ref{fig:0-5NeutMasses} "repel" each
other. The heaviest neutralino smoothly changes its character from the
MSSM wino to the singlino/bino' when $M'_1$ passes the cross-over zone
near 900 GeV. Even more interesting feature occurs when $M'_1$
approaches 2500 GeV, as illustrated in the right panel of
Fig.~\ref{fig:0-5NeutMasses} -- a magnified part of the left
panel. The singlino/bino' state belonging to the negative eigenvalue
(which is now the third state according to mass ordering) mixes
strongly with the second one. It does not mix with the first one since
these states belong to eigenvalues of opposite sign. As a result of
the mixing the mass of the second state is pushed down and below the
lightest one for $M'_1$ above $\sim 2.6$ TeV. Thus the LSP
discontinuously changes its character from being mainly higgsino to
mainly singlino/bino' when $M'_1$ passes the cross-over zone near 2.6
TeV. For higher $M'_1$ values the LSP becomes dominantly
singlino. This behavior will be important to understand
discontinuities in plots to follow.

\subsubsection{Relic density}
\label{scenA-rd}

\begin{figure}[h]
  \begin{center}
    \includegraphics[scale=0.6]{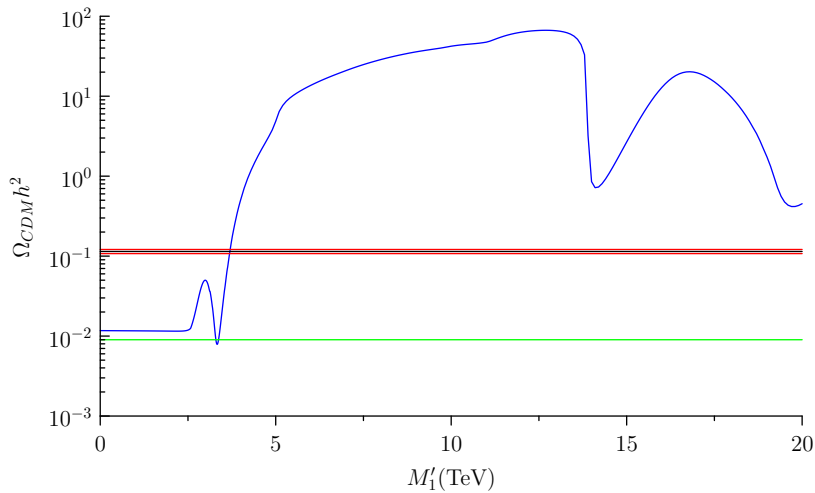}
  \end{center}
  \caption{\emph{The relic density across varying $M_1'$, for
  $\mu=300$~GeV, $m_A=500$~GeV and $\tanb=5$. The red lines show the
  $2\sigma$ measurement of the $\Omega_{CDM}h^2$ by WMAP-5. The green
  line shows the approximate MSSM higgsino relic density for
  $\mu=300$~GeV.}
  \label{fig:0-20Relic}}
\end{figure}

Having set the masses, we vary $M_1'$ and calculate the relic
density. The resulting values for the relic density are plotted in
Fig.~\ref{fig:0-20Relic}. Before dealing with the specific channels
that give rise to the different features, we make some general
points. Firstly, as $\mu=300$~GeV and $m_{A,H,H^\pm}\approx 500$~GeV
it is never possible for a pair of neutralinos to annihilate to a pair
of pseudoscalar Higgs bosons, medium mass Higgs bosons or charged
Higgs bosons in the final state. Secondly, as the squarks and sleptons
are significantly more massive than the mass of the LSP, they do not
contribute significantly to the annihilation cross-section except
where noted below.

In the range $0<M_1'<2.5$~TeV the LSP is predominantly composed of
MSSM-higgsino and gives a relic density of the same order of magnitude
as an MSSM-higgsino. At $M_1=2.57$~TeV the LSP becomes dominantly
singlino, as shown by the cross-over of the mass lines in
Fig.~\ref{fig:0-5NeutMasses}. As $M'_1$ increases, the singlino
component of the LSP increases steadily. This decreases the strength
of the $\tilde{\chi}_1^0-\tilde{\chi}^\pm_1$ coannihilation that
dominates the annihilation amplitude for a predominantly MSSM-higgsino
LSP. As a result we might expect the value of $\Omega_{CDM}h^2$ to
increase noticably before $M_1'=2.57$~TeV. However, as $M'_1$
approaches 2.57~TeV the mass splitting between $\tilde{\chi}_1^0$ and
$\tilde{\chi}_2^0$ decreases. This increases the amplitude for
$\tilde{\chi}_1^0-\tilde{\chi}_2^0$ coannihilation. This increase
compensates the drop in the neutralino-chargino coannihilation and
results in an almost flat value of $\Omega_{CDM}h^2$ up to
$M_1'=2.57$~TeV.

Above $M_1'=2.57$~TeV the mass splitting between the lightest
neutralino and the lightest chargino and next to lightest neutralino
increases steadily. This quickly turns off any coannihilation
processes. At the same time, the singlino component of the lightest
neutralino increases quickly. This steadily reduces the amplitude of
$\tilde{\chi}_1^0-\tilde{\chi}_1^0$ annihilations. As a result of the
combination of these two effects there is a sharp rise in the relic
density above $M_1'=2.57$~TeV.

At $M_1'\approx 3$~TeV we see a sharp dip in the value of
$\Omega_{CDM} h^2$ caused by the pseudoscalar Higgs s-channel
resonance. Just below $M_1'=5$~TeV we see a sharp jump in the relic
density as the LSP drops below the top mass, ruling out processes of
the form $\neut\neut\rightarrow H^*\rightarrow t\overline{t}$. By
$M_1'=5$~TeV the LSP is 94\% singlino with a 3\% bino' admixture and a
2\% higgsino admixture. This, combined with the mass splitting between
the higgsinos and the singlino LSP, suppresses the annihilation of the
singlino resulting in a relic density well above the measured
value. At this point the dominant annihilation channel is to
$b,\overline{b}$ through off-shell s-channel Higgs production, with a
subdominant contribution from t-channel higgsino exchange to final
state light Higgs bosons. A small kink in the relic density profile at
$M_1'\approx 11$~TeV is the point at which the singlino becomes
lighter than the light Higgs boson and final states with two Higgs
bosons become kinematically disallowed. The dip at $M_1'=14$~TeV is
the light Higgs resonance and the dip at $M_1'=20$~TeV is the $Z_1$
resonance.

Here we have seen that the dominant annihilation channels of the
singlino - through t-channel higgsino exchange and through s-channel
Higgs production - are not strong enough to give a relic density in
agreement with the measured value. The exception is when the singlino
is mixed with a higgsino state. This enhances the annihilation through
s-channel Higgs production as the neutralino-neutralino-Higgs vertices
require a non-zero higgsino contribution. It also enhances
annihilation through t-channel higgsino exchange as the higgsinos are
lighter.

The fact that we find a large relic density for a singlino LSP is
partly down to our choice of parameters. Singlino dark matter
dominantly annihilates to Higgs bosons, and with the parameters chosen
above all but the lightest Higgs boson are excluded from the final
state by kinematics and s-channel processes are similarly suppressed
by the large masses. This would not be the case if we were to take
$m_{A,H,H^\pm}<\mu$. We can do this by either lowering $A_\lambda$ or
increasing $\lambda$. Raising $\lambda$ also has the effect of
increasing the coupling strength of the relevant vertices for singlino
annihilation. We will discuss these effects further in scenario B.

\subsubsection{Direct detection}
\label{scenA-dd}

\begin{figure}[h!]
\begin{center}
  \includegraphics[scale=0.5]{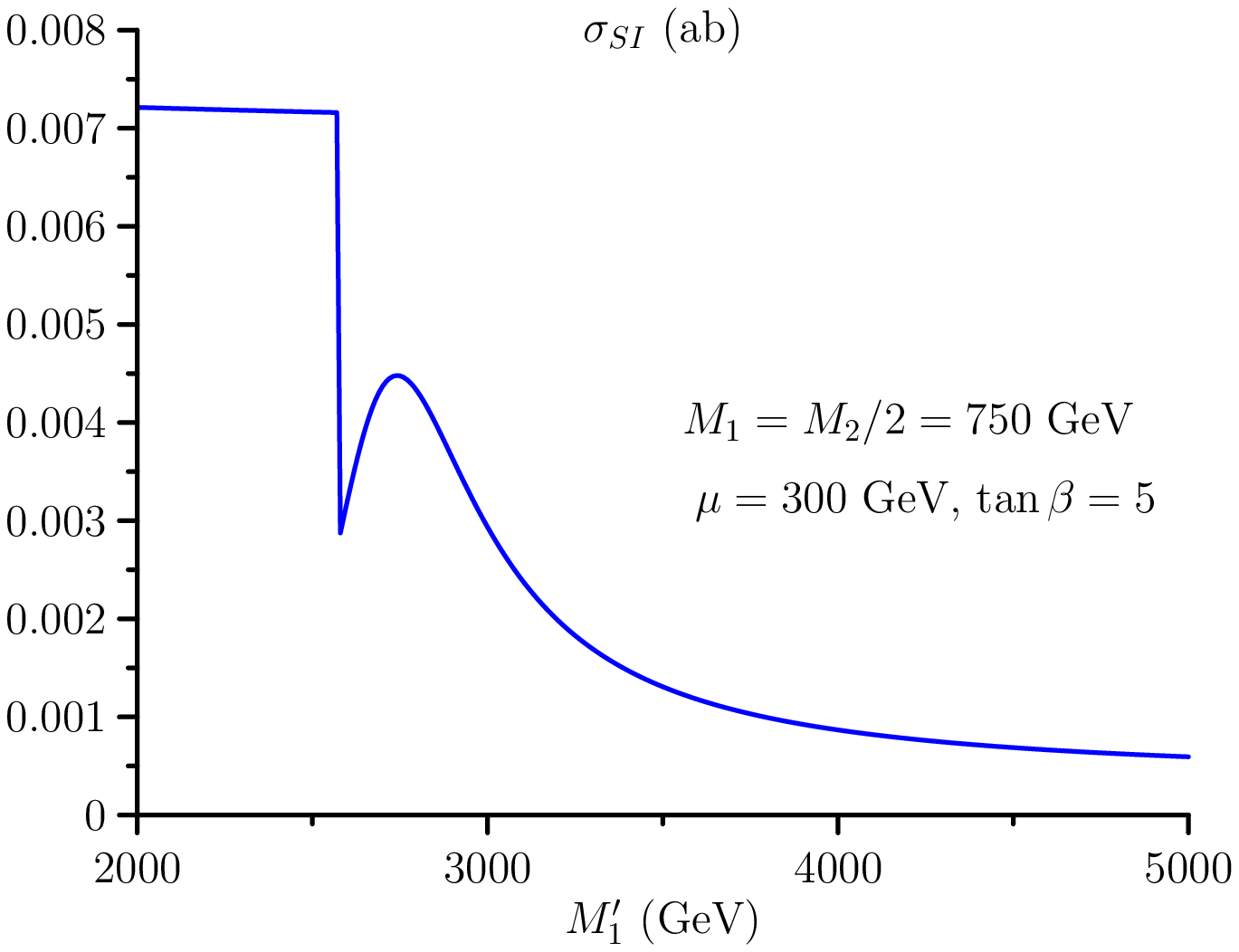}~\includegraphics[scale=0.5]{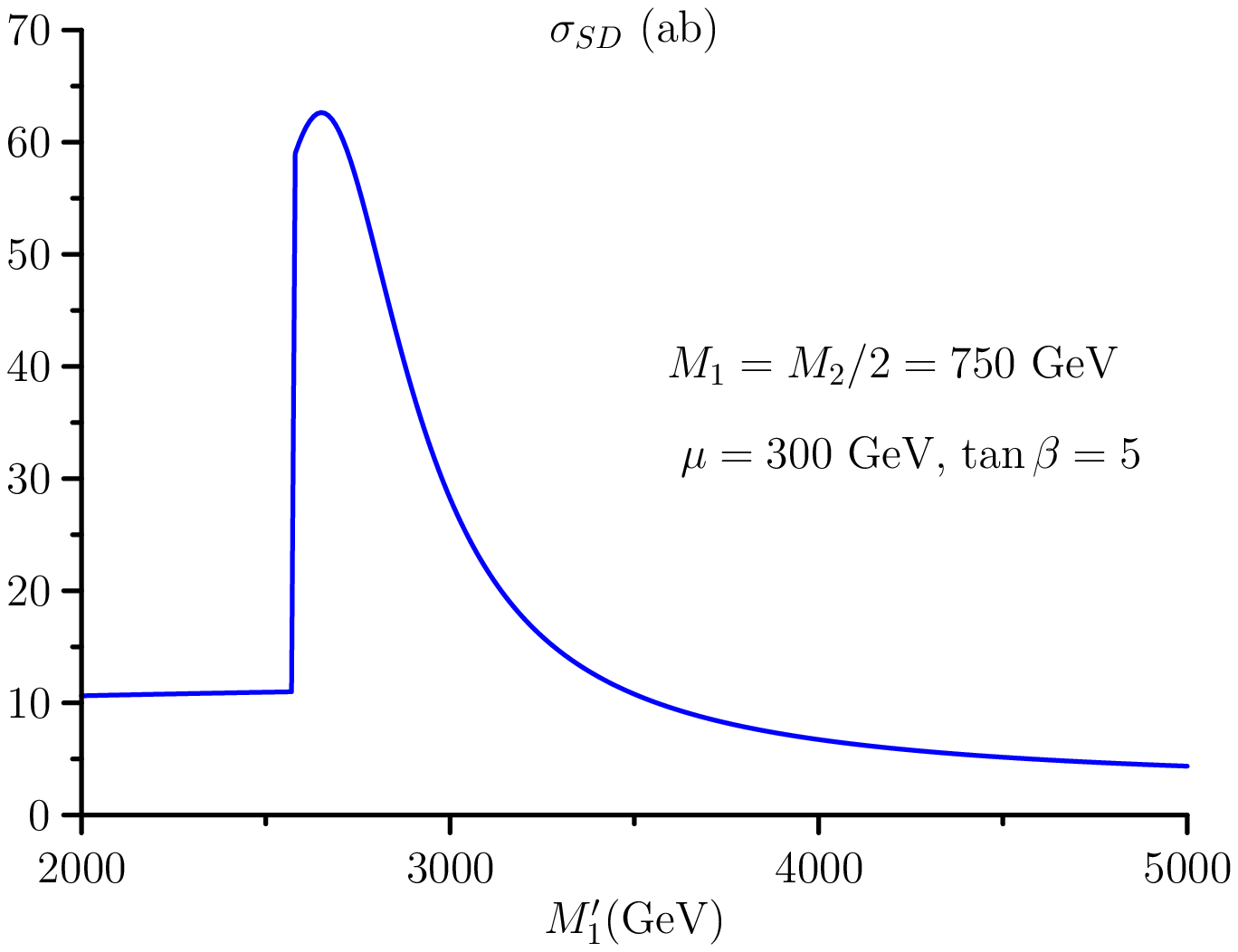}
\end{center}
\caption{\emph{The elastic spin-independent (left) and spin-dependent
    (right) LSP-proton cross section as a function of $M_1'$ in scenario
    A.}\label{fig:non-proton}}
\end{figure}

Let us now turn to the direct DM detection analysis.  In
Fig.~\ref{fig:non-proton} the spin-independent as well as
spin-dependent elastic cross section of the lightest neutralino on a
proton is shown as a function of $M'_1$. We restrict the range of
$M'_1$ to 2--5 TeV, since beyond this range the cross section changes
monotonically.

To understand the $M_1'$ behavior, we refer to
Fig.\ref{fig:0-5NeutMasses}. Up to $M'_{1}\sim 2.5$ TeV the lightest
neutralino is almost a pure MSSM higgsino.  As a result its couplings
do not depend on $M'_{1}$ and the scattering cross sections are
practically determined by the MSSM-like terms $G_k$. Both the SI and
SD cross sections are almost equal to the MSSM result with
corresponding parameters.

The discontinuity in the cross sections around 2.5 TeV is related to
the sudden change of the nature of the LSP. As the $M'_1$ parameter
increases, the mixing between the third and the second states pushes
the latter below the lightest one (right panel of
Fig.\ref{fig:0-5NeutMasses}). The nature of the LSP therefore changes
discontinuously from one of the MSSM-like higgsinos to the other
higgsino state which at the same time acquires an increasing singlino
component.

The reduction of the spin-independent cross section (left panel) can
be understood by realizing that the elastic cross section of the
second-lightest state (according to mass ordering below $M'_1= 2.5$
TeV) on the proton is more than an order of magnitude smaller than
that for the lightest one. When it becomes the LSP (for $M'_1>2.5$
TeV) the SI cross section drops significantly. As the singlino
component of the LSP increases with $M'_1$ the $G_{k}''$ factors,
which are sensitive to both the singlino and the higgsino components
-- viz.~Eq.~(\ref{eq:Gterms}), become responsible for the rise of the
cross section. With further increase of $M'_1$ the LSP becomes almost
a pure singlino which explains a steady fall of the cross section.

The spin-dependent cross section is dominated by the gauge boson
exchange diagram. The $Z$ coupling to the lightest neutralino is
controlled by the combination $c_{34}\equiv |N_{13}|^2-|N_{14}|^2$ of
neutralino mixing matrix elements. For low $M'_1$ the lightest
neutralino is almost a perfect mixture of $\tilde{H}_d^0$ and
$\tilde{H}_u^0$ for which these elements almost entirely cancel
resulting in a small value of $c_{34}$. As $M'_1$ increases the
singlino forces the second-lightest state to become the lightest
(flipping the sign of the coupling) and upsets this delicate
cancelation. As a result, the cross section increases by a factor ~6
and then starts to fall as the LSP becomes dominantly a pure singlino
state.

\subsection{Scenario B: $M_1=M_1'=M_2/2$}

In the previous subsection we have considered the phenomenology of the
USSM with non-universal $M_1$ and $M_1'$. In this section we will
consider the scenario in which gaugino masses are unified at the GUT
scale implying the ratio $M_1:M_1':M_2=1:1:2$ at the electroweak
scale. We will vary $M_1'$ (together with other gaugino masses) as
before and consider the behavior of both the relic density and the
direct detection behavior. Motivated by the remarks at the end of
Subsection~\ref{scenA-rd} we also increase the value of $\mu$
parameter by a factor of 2, i.e. we take $\mu=600$ GeV. This is
achieved by doubling the size of $\lambda$.

\subsubsection{Mass spectrum}
\label{scenB-ms}

\begin{figure}[h!]
\begin{center}
  \includegraphics[scale=0.5]{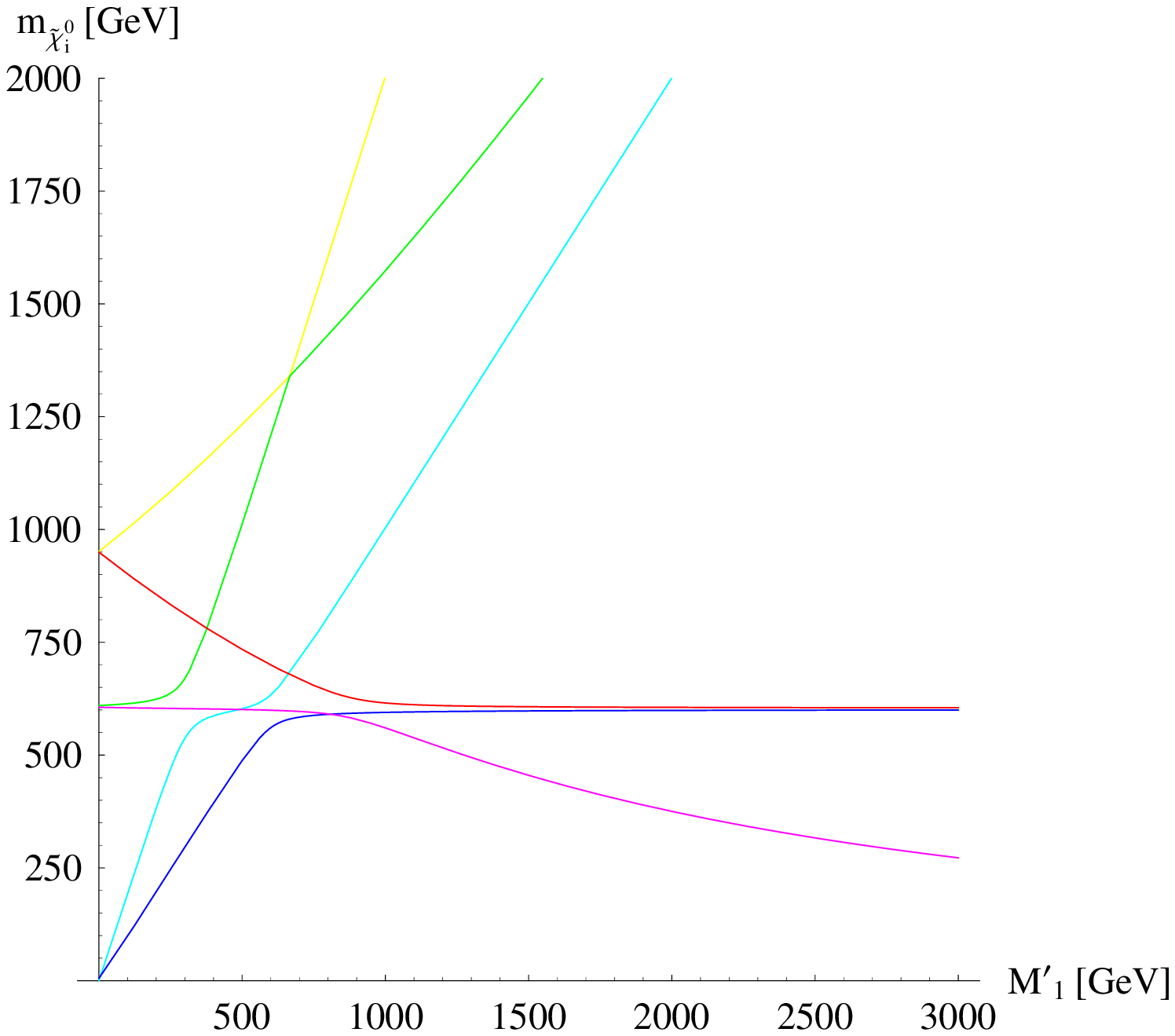}~ \includegraphics[scale=0.5]{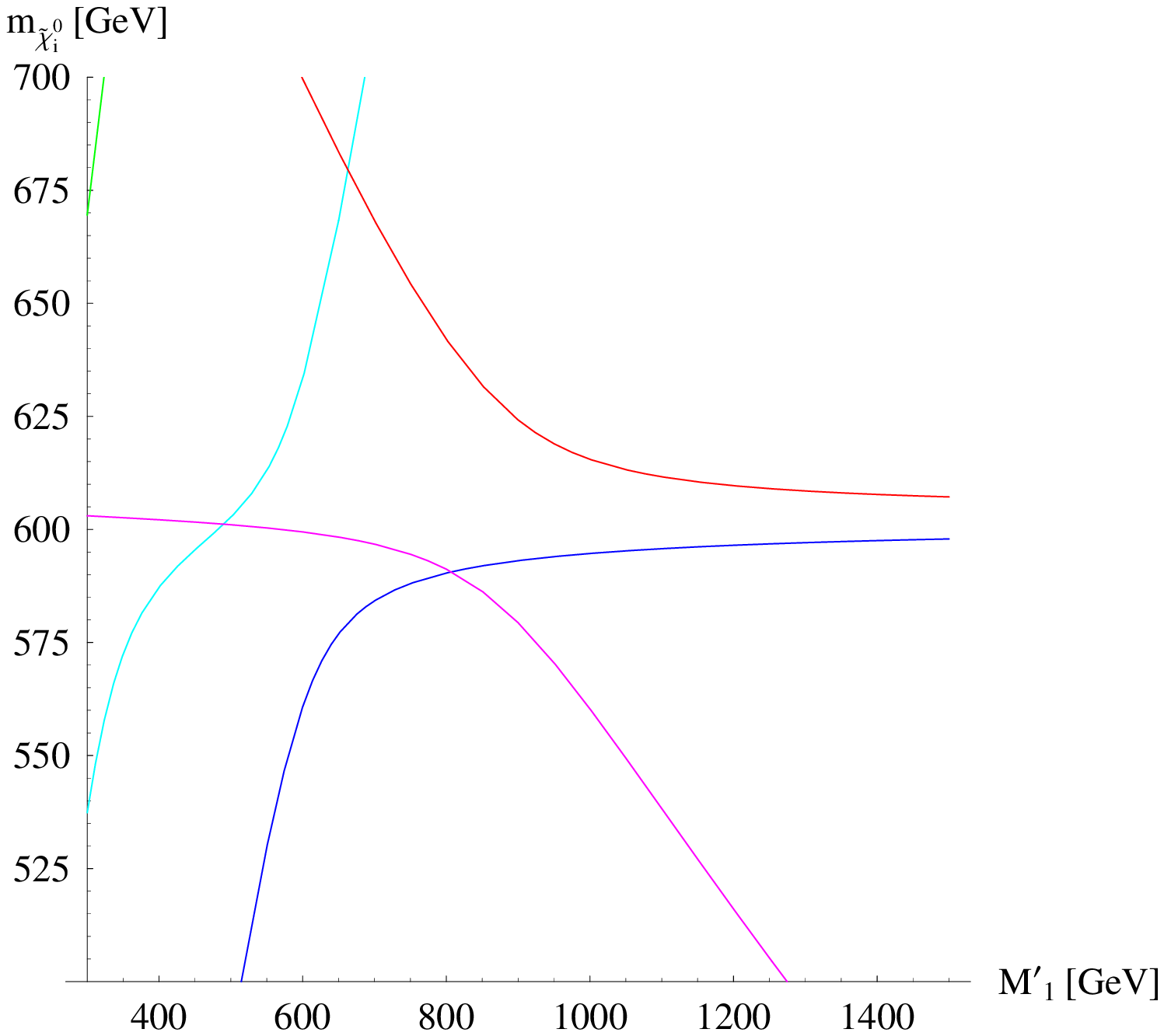}
\end{center}
\caption{\emph{The neutralino mass spectrum as a function of $M_1'$ in
    the unified scenario $M_1'=M_1$. The effective $\mu$ parameter is
    set to 600 GeV, other parameters as in the previous
    subsection. Right panel is a magnified part of the left
    one.}\label{fig:masses-unifiedU1}}
\end{figure}

Again to understand qualitatively the neutralino mixing pattern we
start the discussion with $M'_1=0$. After the Takagi diagonalization
of the neutralino mass matrix at $M'_1=0$ we find two almost massless
eigenstates (dominated by the MSSM bino and wino components), a pair
of nearly degenerate, maximally mixed MSSM higgsinos at $\sim$ 600~GeV
(the third and fourth states) and a pair of nearly degenerate,
maximally mixed singlino/bino' neutralinos at 949~GeV (the fifth and
sixth).  The LEP limit on the lightest chargino mass therefore
enforces $M'_1 \gtrsim 55$ GeV.

For understanding the neutralino mixing pattern as a function of
$M'_1$ it is important to remember that the lighter of the two
singlino/bino' and the lighter of the two higgsino states belong to
negative eigenvalues, while the other states to positive eigenvalues.
When the $M'_1$ parameter is switched on, the mixing pattern is more
rich since not only the singlino/bino', but also the bino and wino
states vary considerably, see Fig.~\ref{fig:masses-unifiedU1}. As a
result there are more cross-over zones where mixing is important. In
the cross-over zone around $M'_1\sim 270$ GeV the wino mixes with the
heavier higgsino, around 500 GeV the bino mixes with the heavier
higgsino, around 550 GeV the wino mixes with the heavier
singlino/bino' and in the last zone around 900 GeV the lighter
higgsino mixes with the lighter singlino/bino' state. This is
illustrated in Fig.~\ref{fig:masses-unifiedU1}.  As the lines develop
from $M'_1=0$ GeV, the dominant component of the corresponding state
changes its nature.  For example, along the green line the state
starts at $M'_1=0$ GeV as a heavier higgsino, then gradually becomes a
wino-dominated (for $M'_1\sim 400-600$ GeV) and finally (for
$M'_1>600$ GeV) a bino'-dominated neutralino. The LSP mass, as we
increase $M_1'$, first increases, then levels off at $M_{LSP}\approx
600$ GeV and then decreases along with $M'_1$.  Its nature also
changes. It starts as a bino, at $M'_1\sim 600$ GeV gradually changes
to a higgsino-dominated state and at $M'_1\sim 800$ GeV
discontinuously jumps to a singlino/bino'-dominated state.  For higher
values of $M'_1$ the lightest neutralino becomes mostly singlino.

\subsubsection{Relic density}
\label{scenB-rd}

\begin{figure}[h]
  \begin{center}
    \includegraphics[scale=0.6]{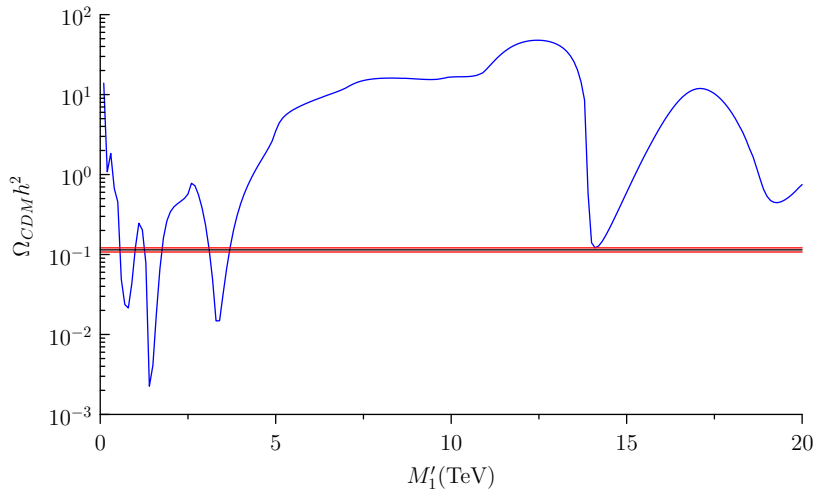}
  \end{center}
  \caption{\emph{The relic density across varying $M_1'$ with
  $M_1'=M_1=M_2/2$, for $\mu=600$~GeV, $m_A=500$~GeV and $\tanb=5$.}
  \label{fig:M1P=M1,Relic}}
\end{figure}

In Fig~(\ref{fig:M1P=M1,Relic}) we show the relic density calculation
for coupled gaugino masses and $\mu=600$~GeV. In this case the relic
density phenomenology is significantly more complex than
previously. First of all, note that below $M_1'=0.75$~TeV the LSP is
predominantly bino, with non-zero admixtures from all other
states. Above $M_1'=0.75$~TeV the LSP is predominantly singlino with
substantial admixtures of bino' and higgsino. Around $M_1'=0.75$~TeV
the LSP is predominantly higgsino with a large admixture of both
singlino and bino.

If we initially ignore the resonances we can see a general trend in
the relic density from a large value at low $M_1'$, down to a lower
value at around $M_1'=0.75$~TeV and then back to larger values at high
$M_1'$. This is to be expected as this follows the evolution of the
LSP from bino (that generally gives $\Omega_{CDM}h^2\gg\Omega_{WMAP}$)
through higgsino (generally $\Omega_{CDM}h^2\ll\Omega_{WMAP}$) to
singlino ($\Omega_{CDM}h^2\gg\Omega_{WMAP}$).

Beyond this general structure there are a number of interesting
features. Note that as $M'_1$ increases the LSP mass first increases
reaching a maximum of $\sim 560$ GeV at $M'_1\sim 800$ GeV and then
falls down crossing all possible $s$-channel resonances
twice. Starting from $M'_1=0$ we first arrive at a little dip in the
relic density around $M'_1=250$ GeV which is due to the $s$-channel
$H_2/A$ resonance. The next resonance due to $Z_2/H_3$ around
$M'_1=500$ GeV produces only a little wiggle since the LSP has not yet
developed an appreciable singlino component.  The first appreciable
dip in the relic density occurs around $M_1'=0.8$~TeV where
$\Omega_{CDM}h^2$ drops to $\sim 0.02$. Here the LSP has a strong
higgsino component which enhances the annihilation via the s-channel
$Z_2/H_3$ resonances considerably. Increasing $M'_1$ further, the LSP
mass increases, going off-resonance (hence local maximum in the relic
density), until it reaches its maximum of $~\sim 590$ GeV at $M'_1\sim
800$ GeV. From now on the LSP mass decreases and its nature becomes
singlino-dominated. Around $M'_1=1.5$ TeV it once again hits the
$Z_2/H_3$ resonance.  However, this time the LSP is predominantly
singlino. Although pure singlino neutralinos do not couple to the
singlet Higgs, so the $H_3$ resonance is subdominant, they couple
strongly to the $Z'$ and annihilate very efficiently.  As a result,
the relic density drops to $\sim 2\times 10^{-3}$.

The next feature of interest is the kink at $M_1'=2.5$~TeV. This is
where the LSP mass drops below threshold for production of $H_1A$
in the final state. This backs up our expectation that annihilation to
heavier Higgs states significantly increases the annihilation rate of
a singlino LSP.

From this point on the relic density profile shows the same essential
features as in Scenario A. We find a pseudoscalar Higgs resonance at
$M_1'=3.5$~TeV, the top threshold at $M_1'=5$~TeV, the light Higgs
threshold at $M_1'=11$~TeV, the light Higgs resonance at $M_1'=14$~TeV
and the $Z$ resonance at $M_1'=19$~TeV. The one important difference
that is worth noting is that in this figure the light Higgs resonance
does lower the relic density to a point where it agrees with the
WMAP-5 measurements. This is due to the doubling of $\lambda$ between
the two cases. This strengthens the coupling of the
singlino-higgsino-higgs vertex.

In our study of Scenario B we can clearly see the effects of
increasing the size of $\mu$. We can have a heavier singlino which
can annihilate to a wider range of final states. The singlino also has
stronger couplings to the other Higgs and higgsino states, further
reducing the relic density. However we see once again that we need to
tune the mass of the singlino through $M_1'$ to fit the relic density,
either through a precise balance of the singlino/higgsino mixture, or
through a careful balance of the singlino mass against the mass of a
boson that mediates annihilation in the s-channel.

\subsubsection{Direct detection}
\label{scenB-dd}

\begin{figure}[h!]
\begin{center}
  \includegraphics[scale=0.5]{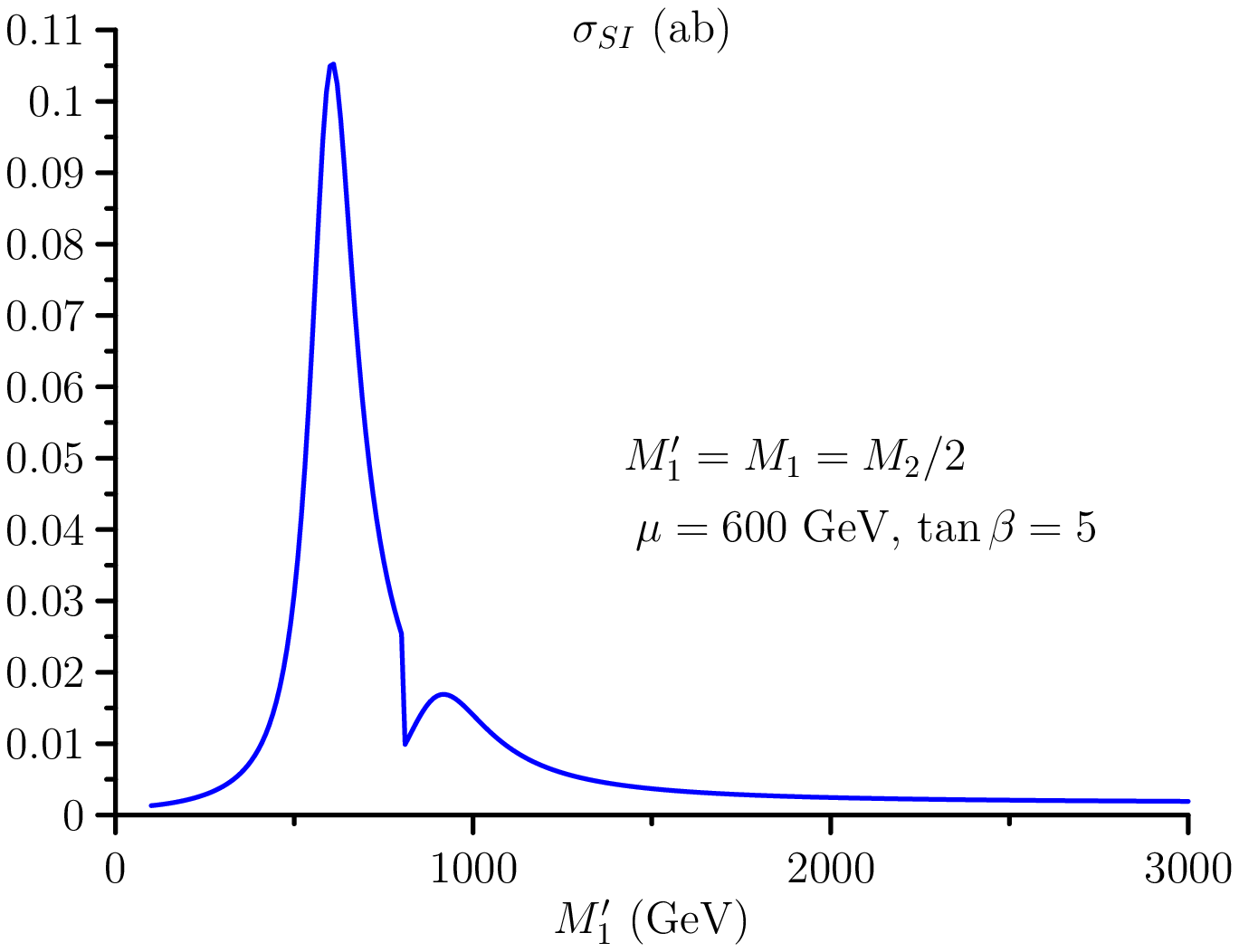}~\includegraphics[scale=0.5]{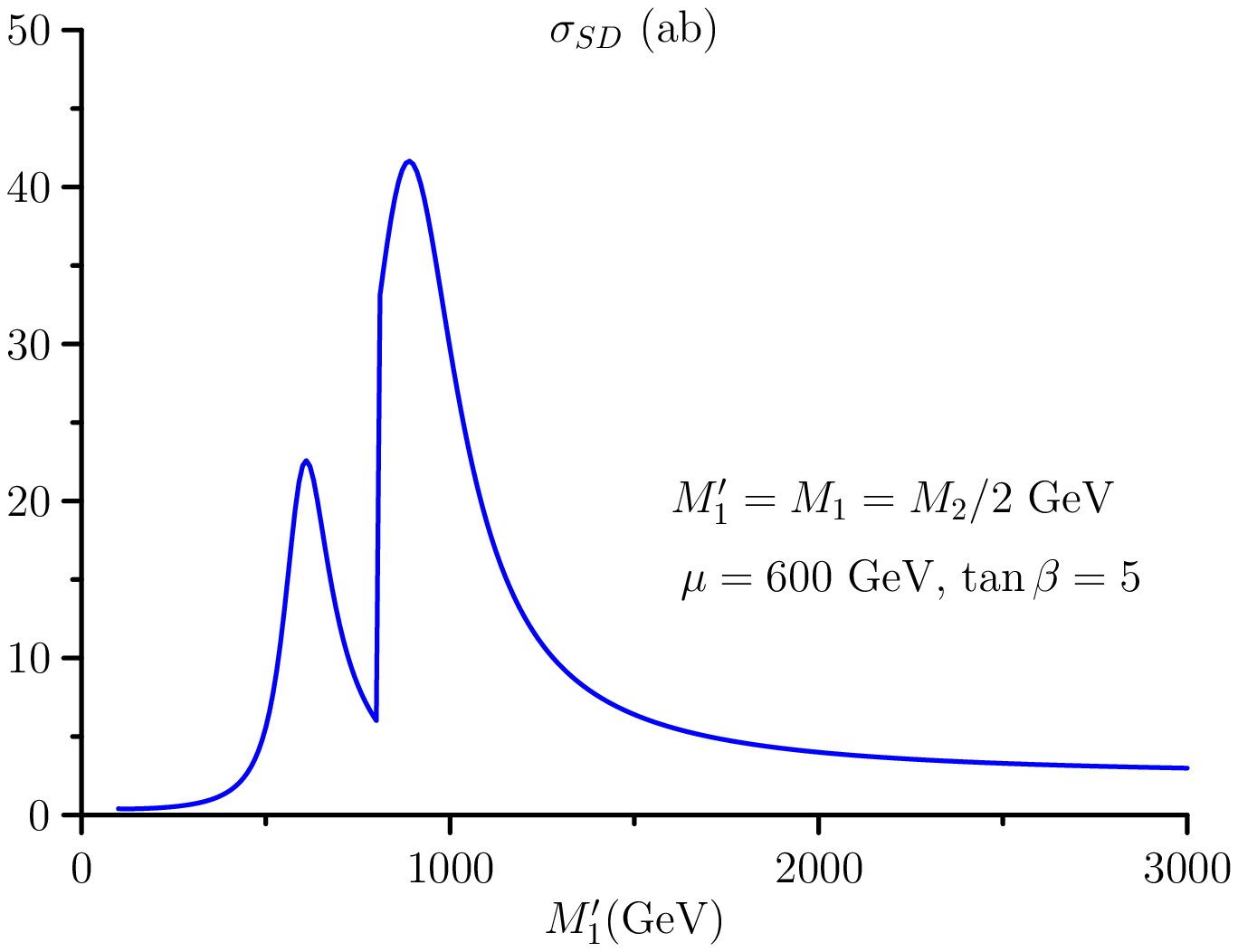}
\end{center}
\caption{\emph{The elastic spin-independent (left) and spin-dependent
  (right) LSP-proton cross section as a function of $M_1'$ in the
  unified scenario $M_1'=M_1$ .}\label{fig:uni-proton}}
\end{figure}

In Fig.~\ref{fig:uni-proton} the spin-independent as well as
spin-dependent elastic cross section of the lightest neutralino on
proton is shown as a function of $M'_1$. We restrict the range of
$M'_1$ to 0--3~TeV, as beyond this range the cross section falls
monotonically.

Referring to Fig.~(\ref{fig:masses-unifiedU1}), it is easy to understand
the $M_1'$ behavior of the cross section. For small $M'_1$ the
lightest neutralino (up to $M'_{1}\sim 0.3$~TeV) is almost a pure MSSM
bino  and its couplings are roughly $M'_{1}$-independent. As $M'_1$
approaches 500 GeV, the LSP receives an appreciable admixture of both
higgsinos. As a result both spin-independent and spin-dependent cross
sections rise. However, the spin-dependent cross section being
sensitive to the combination $c_{34}$ develops a dip around
$M'_1=800$~GeV until the discontinuity where two lightest states
cross. Above 800~GeV the steady increase of the singlino component in
the LSP makes the behavior of the cross section resemble the one in
the previous scenario (for $M'_1>2.5$~TeV).

\section{Summary and conclusions}

In this paper we have provided an up to date and comprehensive
analysis of neutralino dark matter within the USSM which contains, in
addition to the MSSM states, also one additional singlet Higgs plus an
extra $Z'$, together with their superpartners the singlino and
bino'. We have seen that the extra states of the USSM can
significantly modify the nature and properties of neutralino dark
matter relative to that of the MSSM and NMSSM. Using the LanHEP
package, we have derived all the new Feynman rules relevant for the
dark matter calculations. We have also provided a complete qualitative
discussion of the new annihilation channels relevant for the
calculation of the cold dark matter relic density for the neutralino
LSP in the USSM. We also discussed the elastic scattering cross
section for the neutralino LSP in the USSM, including both
spin-independent and spin-dependent parts of the cross sections,
relevant for the direct dark matter search experiments.

We then surveyed the parameter space of the USSM, and discussed
quantitatively how the nature and composition of the neutralino LSP
can be significantly altered compared to that in the MSSM due to the
extra singlino and bino' states, for different ranges of
parameters. We have considered two approaches to the parameter space:
(a) holding the MSSM higgsino and gaugino mass parameters fixed, while
the mass of the extra U(1) gaugino taken free (to complement the
collider phenomenology discussed in Ref.~\cite{Choi:2006fz}); (b) the
scenario of unified gaugino masses.  The Feynman rules were then
implemented into the micrOMEGAs package in order to calculate the
relic density for the corresponding regions of parameter space. This
provides a full calculation of the annihilation channels including
co-annihilation and careful treatment of resonances as well as
accurately calculating the relic density for an arbitrary admixture of
states. In this way we extended the analysis of USSM dark matter
annihilation beyond the specific cases previously studied in the
literature. We also performed an equally general calculation of the
direct detection cross-sections for USSM dark matter for elastic
neutralino--nuclei scattering.

The results show that there are many cases where successful relic
abundances may be achieved, and in novel ways compared to the MSSM
or NMSSM (see for example the low mass region in
Fig.~\ref{fig:M1P=M1,Relic} for $M_1'<5$~TeV). In general our
results also show that the inclusion of the bino' state, as well
as the lack of a cubic interaction term $\hat S^3$, results in a
significant change in the dark matter phenomenology of the USSM as
compared to that of MSSM or NMSSM. Also the neutralino mass
spectrum in the USSM may be very different from that of the NMSSM
as the singlino mass is determined indirectly by a mini-see-saw
mechanism involving the bino' soft mass parameter $M_1'$ rather
than through a diagonal mass term arising from the cubic $\hat
S^3$. The lack of a cubic interaction term also restricts the
annihilation modes of the singlino, making it dominantly reliant
on annihilations involving non-singlet Higgs bosons and higgsinos.
As the USSM has a different Higgs spectrum to the NMSSM, notably
in the pseudoscalar Higgs sector, the Higgs dominated annihilation
channels of the USSM singlino are significantly modified with
respect to the NMSSM singlino. As Higgs exchange diagrams dominate
the direct detection phenomenology, the difference in the Higgs
spectrum and the singlino interactions results in significant
differences in the direct detection predictions as well.

In conclusion, the USSM, despite its modest additional particle
content compared to the MSSM or NMSSM, leads to a surprisingly rich
and interesting dark matter phenomenology which distinguishes it from
these models. The other states which are necessary in order to make
the model anomaly free, and which we have neglected here, can only add
to the richness of the resulting phenomenology, but the qualitatively
new features that we have found in the USSM will remain in any more
complete model. Nevertheless it would be interesting to study the
effect of the additional states present, for example, in the E$_6$SSM
in a future study.

\section*{Acknowledgments}
JK was partially supported by the Polish Ministry of Science and
Higher Education Grant No~1~P03B~108~30. This research was supported
by the EC Programme MTKD--CT--2005--029466 ``Particle Physics and
Cosmology: the Interface'', the EU Network MRTN-CT-2006-035505 ``Tools
and Precision Calculations for Physics Discoveries at Colliders"; STFC
Rolling Grant PPA/G/S/2003/00096; EU Network MRTN-CT-2004-503369; EU
ILIAS RII3-CT-2004-506222; NSF CAREER grant PHY-0449818 and DOE OJI
grant \# DE-FG02-06ER41417. JPR would like to thank Yosi Gelfand and
Neal Weiner for useful discussions. The authors would also like to
thank Dorota Jarecka for collaboration in the early stage of this
work.

\begin{appendix}
\numberwithin{figure}{section} \numberwithin{table}{section}
\numberwithin{equation}{section}

\section{Higgs boson masses}
In general the neutral CP-even Higgs fields $h,H,S$ mix. The mass
matrix takes the form (see the first paper in \cite{King:2005jy})
\begin{equation}
M_{even}^{2}=\left(\begin{array}{ccc}
M_{11}^{2} & M_{12}^{2} & M_{13}^{2}\\
M_{21}^{2} & M_{22}^{2} & M_{23}^{2}\\
M_{31}^{2} & M_{32}^{2} & M_{33}^{2}\end{array}\right)\label{MevenH}\end{equation}
where
\begin{eqnarray}
M_{11}^{2} & = & \frac{\lambda^{2}}{2}v^{2}\sin^{2}2\beta+\frac{g'^{2}+g_{2}^{2}}{4}v^{2}\cos^{2}2\beta+
g_{1}^{'2}v^{2}(Q_{1}\cos^{2}\beta+Q_{2}\sin^{2}\beta)^{2}+\Delta_{11}\,,\nonumber \\
M_{12}^{2} & = & M_{21}^{2}=\left(\frac{\lambda^{2}}{4}-\frac{g'^{2}+g_{2}^{2}}{8}\right)v^{2}\sin4\beta+
\frac{g_{1}^{'2}}{2}v^{2}(Q_{2}-Q_{1})\times\nonumber \\
 &  & \times(Q_{1}\cos^{2}\beta+Q_{2}\sin^{2}\beta)\sin2\beta+\Delta_{12}\,,\nonumber \\
M_{22}^{2} & = & \frac{\sqrt{2}\lambda A_{\lambda}}{\sin2\beta}v_{S}+\left(\frac{g'^{2}+g_{2}^{2}}{4}-
\frac{\lambda^{2}}{2}\right)v^{2}\sin^{2}2\beta+\frac{g_{1}^{'2}}{4}(Q_{2}-
Q_{1})^{2}v^{2}\sin^{2}2\beta+\Delta_{E22}\,,\nonumber \\
M_{13}^{2} & = & M_{31}^{2}=-\frac{\lambda A_{\lambda}}{\sqrt{2}}v\sin2\beta+\lambda^{2}v\, v_{S}+g_{1}^{'2}(Q_{1}\cos^{2}\beta+Q_{2}\sin^{2}\beta)Q_{S}v\, v_{S}+\Delta_{13}\,,\nonumber \\
M_{23}^{2} & = & M_{32}^{2}=-\frac{\lambda A_{\lambda}}{\sqrt{2}}v\cos2\beta+\frac{g_{1}^{'2}}{2}(Q_{2}-Q_{1})Q'_{S}v\, v_{S}\sin2\beta+\Delta_{23}\,,\nonumber \\
M_{33}^{2} & = & \frac{\lambda A_{\lambda}}{2\sqrt{2}v_{S}}v^{2}\sin2\beta+g_{1}^{'2}Q^2_{S}v_{S}^{2}+\Delta_{E33}\,\end{eqnarray}
where the one loop-corrections $\Delta_{Eij}$ are expressed as
\begin{eqnarray}
\Delta_{E22} & = & \Delta_{22}+\Delta_{A}-\Delta_{\beta}\label{deltaE}\\
\Delta_{E33} & = & \Delta_{33}-\frac{\Delta_{S}}{v_{S}}\nonumber \\
\Delta_{EA} & = & \Delta_{A}-\Delta_{\beta}-\frac{\Delta_{S}}{v_{S}}+\Delta_{3}\nonumber \end{eqnarray}
in terms of $\Delta_{ij}$, $\Delta_S$, $\Delta_\beta$, $\Delta_A$ and $\Delta_3$ given explicitly in Ref.~\cite{kovalenko} (note that the expression for $K$ in this paper should read
$K = F-\frac{1}{2}\log\frac{m_{\tilde{t}_{1}}^{2}m_{\tilde{t}_{2}}^{2}}{Q^{4}}$).

\section{Feynman rules}

All the Feynman rules presented here are given in terms of the
interaction of mass eigenstates. As a result the Feynman rules
reference many matrices that rotate from the interaction eigenstate to
the mass eigenstate basis. We briefly summarise them here for ease of
reference:
\begin{itemize}
\item $D_{ij}$ - Z,Z' mixing matrix that transforms from the $Z,Z'$
  eigenstates to the $Z_{1,2}$ mass eigenstates, as defined in
  Eq.~(\ref{eq:Z1Z2})
\item ${\mathcal{O}}'_{ij}$ - Higgs mixing matrix from the mass
  eigenstate basis to the the interaction eigenstate basis, defined in
  Eq.~(\ref{eq:defOp}).
\item $N_{ij}$ - neutralino mixing matrix, defined in Eq.~(\ref{ndef})
\item $U_{ij},V_{ij}$ - standard chargino mixing matrices as in the
  MSSM \cite{Chung:2003fi}.
\item $U^{ij}_{\tilde{f}}$ - squark or slepton mixing matrix, defined
  in Eq.~(\ref{diagM_q}).
\end{itemize}

\begin{figure}
\begin{center}
\scalebox{0.75}{
\begin{picture}(200,150)
\Line(50,25)(100,75)
\Line(100,75)(50,125)
\Photon(100,75)(150,75){2}{4}
\Vertex(100,75){2}
\Text(50,25)[tr]{$\tilde{\chi}_n^0$}
\Text(50,125)[br]{$\tilde{\chi}_l^0$}
\Text(150,75)[l]{$Z_i^\mu$}
\end{picture}}
\end{center}
\caption{The $Z^\mu_i\tilde{\chi}^0_l\tilde{\chi}^0_n$ vertex given in
  Eq.~(\ref{eq:NeutNeutZ})\label{f:NeutNeutZ}}
\end{figure}

Feynman rule for the $Z^\mu_i\tilde{\chi}^0_l\tilde{\chi}^0_n$ vertex
shown in Fig.~\ref{f:NeutNeutZ}:
\begin{eqnarray}
\nonumber &i\gamma_\mu&\left[P_L\left\{
\frac{D_{iZ}g_2}{2\cos\theta_W}
\left(-N_{l3}N_{n3}^*+N_{l4}N_{n4}^*\right)\right.\right.\\
\nonumber&&~~~~~\left.\phantom{\frac{1}{2}}-D_{iZ'}g_1'\left(Q_1N_{l3}N_{n3}^* +
Q_2N_{l4}N_{n4}^* + Q_SN_{l5}N_{n5}^*\right)\right\}\\
\nonumber&&-P_R\left\{\frac{D_{iZ}g_2}{2\cos\theta_W}
\left(-N_{l3}^*N_{n3}+N_{l4}^*N_{n4}\right)\right.\\
&&~~~~~\left.\left.\phantom{\frac{1}{2}}-D_{iZ'}g_1'\left(Q_1N_{l3}^*N_{n3} +
Q_2N_{l4}^*N_{n4} +
Q_SN_{l5}^*N_{n5}\right)\right\}\right]\label{eq:NeutNeutZ}
\end{eqnarray}

\begin{figure}
\begin{center}
\scalebox{0.75}{
\begin{picture}(400,150)
\ArrowLine(50,25)(100,75)
\Line(100,75)(50,125)
\Photon(100,75)(140,75){2}{4}
\ArrowLine(150,75)(140,75)
\Vertex(100,75){2}
\Text(50,25)[tr]{$\tilde{\chi}^+_k$}
\Text(50,125)[br]{$\tilde{\chi}_l^0$}
\Text(150,75)[l]{$W^{-,\mu}$}
\Text(10,75)[l]{(a)}
\ArrowLine(250,25)(300,75)
\Line(300,75)(250,125)
\Photon(300,75)(340,75){2}{4}
\ArrowLine(350,75)(340,75)
\Vertex(300,75){2}
\Text(250,25)[tr]{$\tilde{\chi}^-_k$}
\Text(250,125)[br]{$\tilde{\chi}_l^0$}
\Text(350,75)[l]{$W^{+,\mu}$}
\Text(210,75)[l]{(b)}
\end{picture}}
\end{center}
\caption{The $\tilde{\chi}_k^\pm\tilde{\chi}_l^0W^\mp_\mu$ vertex
  given in (a) Eq.~(\ref{eq:WChargNeuta}) and (b)
  Eq.~(\ref{eq:WChargNeutb})\label{f:WChargNeut}}
\end{figure}

Feynman rules for the $\tilde{\chi}_k^\pm\tilde{\chi}_l^0W^\mp_\mu$
vertex shown in Fig.~\ref{f:WChargNeut}
\begin{equation}
  ig_2\gamma^\mu\left(C_{lk}^LP_L+C_{lk}^RP_R\right)
  \label{eq:WChargNeuta}
\end{equation}
\begin{equation}
ig_2\gamma^\mu\left(C_{lk}^{R*}P_L+C_{lk}^{L*}P_R\right)
  \label{eq:WChargNeutb}
\end{equation}
where
\begin{eqnarray}
C_{lk}^L&=&N_{l2}V_{k1}^*-\frac{1}{\sqrt{2}}N_{l4}V_{k2}^*\\
C_{lk}^R&=&N_{l2}^*U_{k1}-\frac{1}{\sqrt{2}}N_{l3}^*U_{k2}
\end{eqnarray}

\begin{figure}
\begin{center}
\scalebox{0.75}{
\begin{picture}(200,150)
\Line(50,25)(100,75)
\Line(100,75)(50,125)
\DashLine(150,75)(100,75){5}
\Vertex(100,75){2}
\Text(50,25)[tr]{$\tilde{\chi}_n^0$}
\Text(50,125)[br]{$\tilde{\chi}_l^0$}
\Text(150,75)[l]{$H_k$}
\end{picture}}
\end{center}
\caption{The $\tilde{\chi}_l^0\tilde{\chi}_n^0H_k$ vertex given in
  Eq.~(\ref{hxx})\label{f:hxx}}
\end{figure}

Feynman rule for the $\tilde{\chi}_l^0\tilde{\chi}_n^0H_k$ vertex
shown in Fig.~\ref{hxx}:
\begin{equation}
i\left({\cal O}'_{1k}R_{ln}^* + {\cal O}'_{2k}S^*_{ln} + {\cal
O}'_{3k}T^*_{ln}\right)P_L +\left({\cal O}'_{1k}R_{nl} + {\cal
O}'_{2k}S_{nl} + {\cal O}'_{3k}T_{nl}\right)P_R
\label{hxx}
\end{equation}
where
\begin{eqnarray}
\nonumber R_{nl}&=&-\frac{g_2}{2}(N_{n2}-\tan\theta_WN_{n1})N_{l3} -
g_1'Q_1N_{n6}N_{l3} +
\frac{\lambda}{\sqrt{2}}N_{n4}N_{l5}\\&&+(l\leftrightarrow n)\\
\nonumber S_{nl}&=&\frac{g_2}{2}(N_{n2}-\tan\theta_WN_{n1})N_{l4} -
g_1'Q_2N_{n6}N_{l4} +
\frac{\lambda}{\sqrt{2}}N_{n3}N_{l5}\\&&+(l\leftrightarrow n)\\
T_{nl}&=&-g_1'Q_SN_{n6}N_{l5} + \frac{\lambda}{\sqrt{2}}N_{n3}N_{l4} +
(l\leftrightarrow n)
\end{eqnarray}

\begin{figure}
\begin{center}
\scalebox{0.75}{
\begin{picture}(200,150)
\Line(50,25)(100,75)
\Line(100,75)(50,125)
\DashLine(150,75)(100,75){5}
\Vertex(100,75){2}
\Text(50,25)[tr]{$\tilde{\chi}_n^0$}
\Text(50,125)[br]{$\tilde{\chi}_l^0$}
\Text(150,75)[l]{$A$}
\end{picture}}
\end{center}
\caption{The $\tilde{\chi}_l^0\tilde{\chi}_n^0A$ vertex given in
  Eq.~(\ref{Axx})\label{f:Axx}}
\end{figure}

Feynman rule for the $\tilde{\chi}_l^0\tilde{\chi}_n^0A$ vertex shown
in Fig.~\ref{Axx}:
\begin{eqnarray}
\nonumber &&[(R_{ln}^{'*}\sin\beta+S_{ln}^{'*}\cos\beta)\cos\phi +
T_{ln}^{'*}\sin\phi]P_L\\
&-&[(R'_{nl}\sin\beta+S'_{nl}\cos\beta)\cos\phi +
T'_{nl}\sin\phi]P_R
\label{Axx}
\end{eqnarray}
where
\begin{eqnarray}
\nonumber R'_{nl}&=&
-\frac{g_2}{2}\left(N_{l2}-\tan\theta_WN_{l1}\right)N_{n3} -
g_1'Q_1N_{l3}N_{n6} -
\frac{\lambda}{\sqrt{2}}N_{l4}N_{n5}\\&&+(l\leftrightarrow n)\\
\nonumber S'_{nl}&=&
\frac{g_2}{2}\left(N_{l2}-\tan\theta_WN_{l1}\right)N_{n4} -
g_1'Q_2N_{l4}N_{n6} - \frac{\lambda}{\sqrt{2}}N_{l3}N_{n5}\\&& +
(l\leftrightarrow n)\\
T'_{nl}&=&-g_1'Q_SN_{l5}N_{n6}-\frac{\lambda}{\sqrt{2}}N_{l3}N_{n4}
\end{eqnarray}

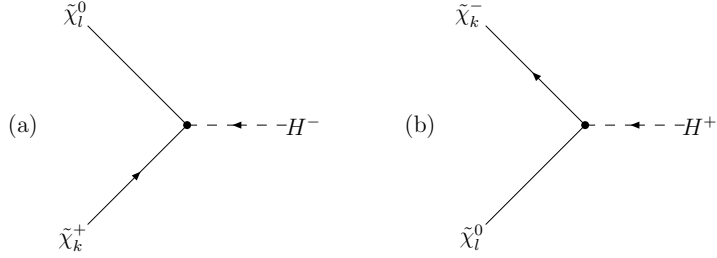
\begin{figure}
\begin{center}
\scalebox{0.75}{
\begin{picture}(400,150)
\ArrowLine(50,25)(100,75)
\Line(100,75)(50,125)
\DashArrowLine(150,75)(100,75){5}
\Vertex(100,75){2}
\Text(50,25)[tr]{$\tilde{\chi}^+_k$}
\Text(50,125)[br]{$\tilde{\chi}_l^0$}
\Text(150,75)[l]{$H^-$}
\Text(10,75)[l]{(a)}
\Line(250,25)(300,75)
\ArrowLine(300,75)(250,125)
\DashArrowLine(350,75)(300,75){5}
\Vertex(300,75){2}
\Text(250,25)[tr]{$\tilde{\chi}^0_l$}
\Text(250,125)[br]{$\tilde{\chi}_k^-$}
\Text(350,75)[l]{$H^+$}
\Text(210,75)[l]{(b)}
\end{picture}}
\end{center}
\caption{The $\tilde{\chi}^\pm_k \tilde{\chi}^0_l H^\mp$ vertex
  given in (a) Eq.~(\ref{xx+H-}) and (b)
  Eq.~(\ref{xx-H+})\label{f:HChargNeut}}
\end{figure}

Feynman rules for $\tilde{\chi}^\pm_k \tilde{\chi}^0_l H^\mp$ shown in
Fig.~\ref{f:HChargNeut}:
\begin{equation}
-i\left(R^{''L}_{lk}P_L+R_{lk}^{''R}P_R\right)
\label{xx+H-}
\end{equation}
\begin{equation}
-i\left(R^{''R*}_{lk}P_L+R_{lk}^{''L*}P_R\right)
\label{xx-H+}
\end{equation}
where
\begin{eqnarray}
\nonumber R_{lk}^{''L}&=&g_2\cos\beta\left[N_{l4}^*V_{k1}^* +
\frac{V_{k2}^*}{\sqrt{2}}
\left(N_{l2}^*+N_{l1}^*\tan\theta_W\right)\right]\\
&&+g_1'\sqrt{2}\cos\beta Q_2N_{l6}^*V_{k2}^* + \lambda\sin\beta
N_{l5}^*V_{k2}^*\\
\nonumber R_{lk}^{''R}&=&g_2\sin\beta\left[N_{l3}U_{k1} -
\frac{U_{k2}}{\sqrt{2}}
\left(N_{l2}-N_{l1}\tan\theta_W\right)\right]\\
&&+g_1'\sqrt{2}\sin\beta Q_1N_{l6}U_{k2} + \lambda\cos\beta
N_{l5}U_{k2}
\end{eqnarray}

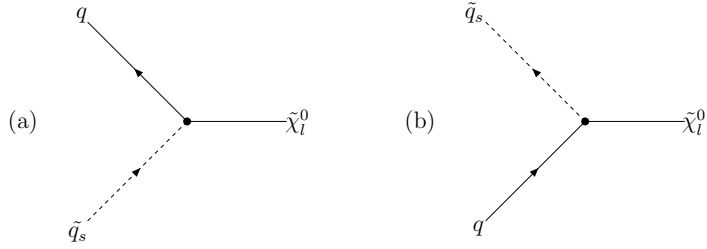
\begin{figure}
\begin{center}
\scalebox{0.75}{
\begin{picture}(400,150)
\DashArrowLine(50,25)(100,75){2}
\ArrowLine(100,75)(50,125)
\Line(150,75)(100,75)
\Vertex(100,75){2}
\Text(50,25)[tr]{$\tilde{q_s}$}
\Text(50,125)[br]{$q$}
\Text(150,75)[l]{$\tilde{\chi}^0_l$}
\Text(10,75)[l]{(a)}
\ArrowLine(250,25)(300,75)
\DashArrowLine(300,75)(250,125){2}
\Line(350,75)(300,75)
\Vertex(300,75){2}
\Text(250,25)[tr]{$q$}
\Text(250,125)[br]{$\tilde{q}_s$}
\Text(350,75)[l]{$\tilde{\chi}_l^0$}
\Text(210,75)[l]{(b)}
\end{picture}}
\end{center}
\caption{The $q \tilde{q}_s  \tilde{\chi}^0_l$ vertex
  given in (a) Eq.~(\ref{xSqQa}) and (b)
  Eq.~(\ref{xSqQb})\label{f:xSqQ}}
\end{figure}

Feynman rules for the $q \tilde{q}_s \tilde{\chi}^0_l$ vertex shown in
Fig.~\ref{f:xSqQ}:
\begin{equation}
i\left[(G_{sl}^{q_L})^*P_R+(G_{sl}^{q_R})^*P_L\right]\label{xSqQa}
\end{equation}
\begin{equation}
i\left[G_{sl}^{q_L}P_L+G_{sl}^{q_R}P_R\right]\label{xSqQb}
\end{equation}

For up-type quarks:
\begin{eqnarray}
\nonumber G_{sl}^{u_L}&=&-\sqrt{2}\left[g_2\left(\frac{1}{2}N_{l2}^* +
\frac{1}{6}\tan\theta_WN_{l1}^*\right) +
g_1'Q_QN_{l6}^*\right]U_{\tilde{u}_i}^{1s} -
\frac{g_2m_{u_i}}{\sqrt{2}M_W\sin\beta}N_{l4}^*U_{\tilde{u}_i}^{2s}\\
G_{sl}^{u_R}&=&\sqrt{2}\left[g_2\frac{2}{3}\tan\theta_WN_{l1} -
g_1'Q_{\bar u}N_{l6}\right]U_{\tilde{u}_i}^{2s} -
\frac{g_2m_{u_i}}{\sqrt{2}M_W\sin\beta}N_{l4}U_{\tilde{u}_i}^{1s}
\label{eq:binoup}
\end{eqnarray}

For down-type quarks:
\begin{eqnarray}
\nonumber G_{sl}^{d_L}&=&\sqrt{2}\left[g_2\left(\frac{1}{2}N_{l2}^* -
\frac{1}{6}\tan\theta_WN_{l1}^*\right) -
g_1'Q_QN_{l6}^*\right]U_{\tilde{d}_i}^{1s} -
\frac{g_2m_{d_i}}{\sqrt{2}M_W\cos\beta}N_{l3}^*U_{\tilde{d}_i}^{2s}\\
G_{sl}^{d_R}&=&-\sqrt{2}\left[\frac{1}{6}\tan\theta_WN_{l1} +
g_1'Q_{\bar d}N_{l6}\right]U_{\tilde{d}_i}^{2s} -
\frac{g_2m_{d_i}}{\sqrt{2}M_W\cos\beta}N_{l3}U_{\tilde{d}_i}^{1s}
\label{eq:binodown}
\end{eqnarray}

\end{appendix}

\end{document}